\documentclass[letterpaper,twocolumn,10pt]{article}
\usepackage{usenix}

\usepackage{tikz}
\usepackage{amsmath}
\usepackage{subfigure}
\usepackage{booktabs}
\usepackage{url}
\usepackage[ruled,linesnumbered]{algorithm2e}
\usepackage{amsmath}
\usepackage{amssymb}
\usepackage{color}
\usepackage{tabularx}
\usepackage{multirow}
\usepackage{CJKutf8}
\usepackage{graphicx}
\usepackage{xcolor}
\usepackage{colortbl}

\definecolor{revision_color}{rgb}{0.0, 0.0, 0.0}

\begin{document}

\date{}

\title{\Large \bf KnowPhish: Large Language Models Meet Multimodal Knowledge Graphs for Enhancing Reference-Based Phishing Detection}
\author{
{\rm Yuexin Li\textsuperscript{1}, 
Chengyu Huang\textsuperscript{1}, 
Shumin Deng\textsuperscript{1}, 
Mei Lin Lock\textsuperscript{2}, 
Tri Cao\textsuperscript{1}},\\
{\rm Nay Oo\textsuperscript{2}, 
Hoon Wei Lim\textsuperscript{2},
Bryan Hooi\textsuperscript{1}}\\
\normalsize \textsuperscript{\rm 1}\textit{National University of Singapore}, \textsuperscript{\rm 2}\textit{NCS Cyber Special Ops-R\&D}\\
} 

\maketitle

\begin{abstract}

    Phishing attacks have inflicted substantial losses on individuals and businesses alike, necessitating the development of robust and efficient automated phishing detection approaches. Reference-based phishing detectors (RBPDs), which compare the logos on a target webpage to a known set of logos, have emerged as the state-of-the-art approach. However, a major limitation of existing RBPDs is that they rely on a manually constructed brand knowledge base, making it infeasible to scale to a large number of brands, which results in false negative errors due to the insufficient brand coverage of the knowledge base. To address this issue, we propose an automated knowledge collection pipeline, using which we collect 
    a large-scale multimodal brand knowledge base, KnowPhish, containing 20k brands with rich information about each brand. KnowPhish can be used to boost the performance of existing RBPDs in a plug-and-play manner. A second limitation of existing RBPDs is that they solely rely on the image modality, ignoring useful textual information present in the webpage HTML. To utilize this textual information, we propose a Large Language Model (LLM)-based approach to extract brand information of webpages from text. Our resulting multimodal phishing detection approach, KnowPhish Detector (KPD), can detect phishing webpages with or without logos. We evaluate KnowPhish and KPD on a manually validated dataset, and a field study under Singapore's local context, showing substantial improvements in effectiveness and efficiency compared to state-of-the-art baselines.    
\end{abstract}

\section{Introduction}
\label{sec:introduction}

Phishing attacks are one of the most impactful types of scams, harming both individuals and businesses: in 2023, an estimated \$1.026 trillion was lost by consumers worldwide in scams~\cite{gasa}, of which phishing scams are among the most common types. Phishing scams also account for about 90\% of data breaches for organizations~\cite{cisco}. This problem has been exacerbated by the proliferation of automated phishing kits, enabling malicious actors to easily mimic genuine pages while evading detection using countermeasures~\cite{zieni2023phishing}. Accordingly, the number of phishing attacks has increased by 47.2\% in 2022 compared to the previous year~\cite{zscaler}. These underscore the importance and urgency of tackling the issue, and the need for effective automatic phishing detection approaches. 


Many efforts have been made to counter phishing attacks, including anti-phishing blacklists, heuristic-based, and feature-based models \cite{phishing_review}. Blacklist-based methods~\cite{openphish, phishtank, google_safe_browsing} compare an input URL with a predefined blacklist of malicious URLs, but these are reactive approaches that require phishing sites to be first reported or detected through other means. Heuristic-based~\cite{shlr} and feature-based models~\cite{urlnet, urltran, hinphish, cantina+, stackmodel, url_1, html_url_1} extract features to proactively identify new phishing webpages. However, as these do not utilize logo information, they are greatly limited in their ability to detect plentiful phishing pages which are mainly identifiable by the presence of a logo~\cite{phishintention}. Moreover, as they depend on statistical features for detecting phishing pages, they are susceptible to distribution shifts due to the constantly changing nature of phishing campaigns.

In contrast, reference-based phishing detectors (RBPDs), which work by comparing the logos on a target webpage to a known set of logos, have been established as the state-of-the-art phishing detection paradigm, garnering considerable research attention~\cite{visualphishnet, phishpedia, phishintention}. Specifically, an RBPD consists of \emph{brand knowledge base} (BKB) containing brand information (the logos and legitimate domains of brands) and a \emph{detector backbone} which uses information from this BKB for phishing detection. To detect if a webpage is a phishing or benign page, RBPDs first identify the webpage's \emph{brand intention}, i.e., the brand that the webpage presents itself as (e.g., a webpage with an Adobe logo and appearance has the brand intention of Adobe). Then, if the webpage is detected to have an intention of a certain brand, but its domain does not match the legitimate domains of that brand, the webpage is classified as phishing. As virtually all phishing pages do not utilize the legitimate domains of the original brand, this approach typically obtains high precision. Moreover, as this approach is based on an invariant that does not change over time, it is relatively robust to distribution shift~\cite{phishpedia, phishintention}.


\noindent\textbf{Challenges.}\ \ 
Despite the advantages of RBPDs, they have two major limitations which we focus on. 
\textbf{(1)} The first limitation is \emph{Limited-Scale BKB}: RBPDs fundamentally rely on their BKB to identify the brand intention of a website. However, it is labor-intensive to construct and maintain a large-scale BKB manually. Phishpedia~\cite{phishpedia} and PhishIntention~\cite{phishintention} rely on manual curation, hence are limited to a small BKB of 277 brands. A recent method, DynaPhish~\cite{dynaphish}, proposes to dynamically expand the BKB during deployment time. However, this leads to extremely long running time, e.g., averaging $10.6$ seconds per sample in our experiments. 
\textcolor{revision_color}{\hypertarget{target:challenges}{}
Moreover, it may fail to construct brand knowledge of novel phishing targets if phishing pages' logos are different from those displayed on legitimate pages.
} 
\textbf{(2)} The second limitation is \emph{Textual Brand Intention}: Phishing webpages can convey their brand intention via text in HTML, instead of via logos. Existing RBPDs cannot identify it because they solely operate within the image modality.

\noindent \textbf{Present Work.}\ \  
In this paper, we seek to address both of these issues \textcolor{revision_color}{in the context of a static environment, where input webpages are not interactable}. First, we propose an automated knowledge collection pipeline, with which we construct 
a large-scale multimodal BKB named \textit{KnowPhish}. KnowPhish is constructed based on our empirical analysis which finds that phishing targets mostly belong to a few high-value industries. Hence, using brand-industry relations modeled from a publicly available knowledge base, Wikidata\cite{wikidata}, we search for a set of potential phishing targets and their brand knowledge predictively, leading to a BKB covering around 20k potential phishing targets. \textcolor{revision_color}{\hypertarget{target:present-work}{}
We also incorporate extra data sources such as Tranco domain list\cite{tranco} and Google Image Search\cite{google_images_search} for brand knowledge enhancement.} Therefore, KnowPhish is equipped with rich logo, alias, and domain variants of each brand, which can be used to boost the performance of existing RBPDs in a plug-and-play manner.

Next, to address the issue of textual brand intention, we develop a Large Language Model (LLM)-based approach to identify the brand intention of webpages in conjunction with the alias information in our KnowPhish BKB. Our approach can be directly integrated with any standard visual RBPD~\cite{phishintention}, augmenting it to form a \emph{multimodal RBPD} which detects brand intention through both visual and textual modalities. Our resulting multimodal phishing detection approach, named \textit{KnowPhish Detector} (\textit{KPD}), \textcolor{revision_color}{can operate with static webpage data to }detect phishing webpages with or without logos. 

We then evaluate the effectiveness and efficiency of KnowPhish and KPD on \texttt{TR-OP}, a manually validated dataset that comprises 5k benign and 5k phishing webpages. We also evaluate our approach on a field study under Singapore's local context, to study how well different approaches generalize to such a local context. The resulting data, which we call \texttt{SG-SCAN}, contains 10k webpages from Singapore local webpage traffic over 6 months. In experiments on the two datasets, KnowPhish significantly boosts the effectiveness of various RBPDs, and is 30 or more times faster than the on-deployment framework DynaPhish~\cite{dynaphish}, when equipped with image-based RBPDs. Moreover, incorporating our multimodal approach, KPD, can substantially boost the number of detected phishing webpages.

In summary, our contributions are three-fold:
\begin{itemize}
    \item \textbf{Multimodal Brand Knowledge Base.} We propose KnowPhish, a large-scale multimodal BKB for phishing detection, and its automated construction approach. KnowPhish can be used in any RBPDs to boost their brand knowledge in a plug-and-play manner.
    
    \item \textbf{Multimodal Reference-based Phishing Detector.} We propose an LLM-based approach to identify textual brands from HTML to handle logo-less phishing webpages. Our approach directly integrates with any existing RBPD, augmenting it to form a multimodal RBPD,  \textcolor{revision_color}{named KnowPhish Detector (KPD), which can detect phishing webpages with or without logos}.
    
    \item \textbf{Effectiveness and Efficiency} Extensive experiments show that \textcolor{revision_color}{KnowPhish significantly enhances the performance of various RBPDs, including KPD, \hypertarget{target:contribution}{}while achieving much better runtime efficiency than DynaPhish.} We also demonstrate their effectiveness in a field study and validate the robustness \textcolor{revision_color}{of KPD} to adversarial attacks. 
\end{itemize}
{\color{revision_color}{
\section{Motivating Examples}
\label{sec:motivating_examples}

The development of KnowPhish and KPD is motivated by the brand knowledge construction difficulty of DynaPhish. Below, we present a brief workflow of DynaPhish, a few examples of where DynaPhish fails to build brand knowledge, and how KnowPhish and KPD address these difficulties.

DynaPhish expands brand knowledge through a two-step process when encountering a webpage with an unseen logo. (1) It first utilizes Google Search (GS) to validate the webpage's domain popularity to determine its benignity. The logo-domain pair will be added to its brand knowledge if such validation passes. (2.1) If the validation fails, it recognizes the brand name of the logo using Google Logo Detector (GLD) and searches for the legitimate domain associated with that logo using another GS with the brand name as the query. (2.2) It then compares the input logo with the logo on each webpage from the GS results. If a match is found, the corresponding domain and logo are added to the brand knowledge.

However, DynaPhish may fail to construct new brand knowledge and correctly classify phishing pages due to the failure in step (2.2), i.e., logo-matching. A few examples of logos from phishing pages and the logos from their corresponding GS result pages are shown in \hyperref[fig:comparison]{Figure \ref{fig:comparison}}. The difference in appearance between the logos on phishing pages and GS result pages causes failure in the logo-matching process, even if GLD identifies the correct brand names, further resulting in brand knowledge construction failure and false negatives.



\begin{figure}
    \centering
    \includegraphics[scale=0.4]{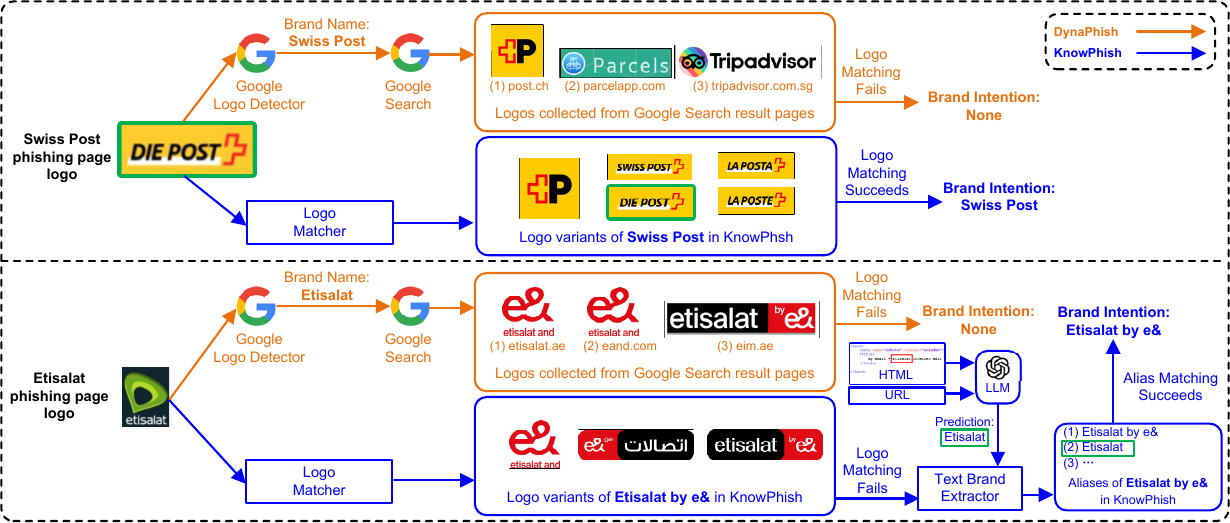}
    \caption{Comparison of the workflow between DynaPhish and KnowPhish to identify the brand intention of the two phishing page examples.}
    \label{fig:comparison}
\end{figure}

These examples motivate us to address such false negative problems in two ways. Firstly, we build KnowPhish from a high-quality data source, Wikidata, based on brand profiles (e.g., names and domains). This approach bypasses the logo-matching process used by DynaPhish and directly incorporates enhanced brand knowledge, such as logos and aliases. Therefore, the enriched logo variants increase the chance for image-based RBPDs to match the correct brand. For instance, the Swiss Post phishing page logo in \hyperref[fig:comparison]{Figure \ref{fig:comparison}} is covered in KnowPhish, thus enabling the logo-matcher to identify its brand intention. Secondly, even if image-based RBPDs fail to match the logo, we remedy this by also using the text-brand extractor from KPD to identify the text brand intention from HTML and URL, in conjunction with the rich aliases in KnowPhish. As the Etisalat phishing page in \hyperref[fig:comparison]{Figure \ref{fig:comparison}} shows, an LLM takes both the HTML and URL as input and outputs a predicted brand for the page. The predicted brand matches an alias from KnowPhish, thus the text-brand extractor detects its brand intention as the brand of that alias. We will elaborate on each in Sections \ref{sec:knowphish_construction} and \ref{sec:knowphish_detector}, respectively.



}}

\section{Formalization}
\label{sec:formalization}
\noindent\textbf{Threat Model.}\ \ 
In a phishing attack, a malicious actor misleads visitors into believing that their webpage comes from a legitimate brand, misleading users into providing their credentials (such as username and password). Note that we do not consider other types of attacks, such as malware or miscellaneous scams that do not fit the above description, to be within our scope.

In our threat model, we assume the attacker has full control over their webpage. However, to maintain their webpage's effectiveness in misleading users, the webpage needs to convey its \emph{brand intention}, i.e., present itself as a particular recognizable brand to visitors. In addition, the webpage needs to be \emph{credential receiving} in some way: e.g., via username and password fields, but less common approaches exist such as buttons or QR codes. Deviating from either of these two conditions makes it less likely for the phishing attack to succeed, and our empirical observations support that these two conditions hold consistently in phishing webpages.

Formally, consider a webpage $w$, which consists of its screenshot (\textit{w.screenshot}), its page HTML (\textit{w.html}), and its domain (\textit{w.domain}). As described above, the webpage needs to convey its brand intention, presenting itself as belonging to a brand $b$. This can be done either in visual form (e.g., through logos in its screenshot) or textual form (e.g., through text in its HTML), so brand information may appear either in $\textit{w.screenshot}$ or $\textit{w.html}$ (or both).

\noindent\textbf{Reference-Based Phishing Detection.}\ \ 
Since phishing webpages need to convey brand intention, the state-of-the-art RBPD approach relies on identifying this brand intention, by comparing images on the page to a set of known \emph{reference} logos. Formally, an RBPD consists of a \emph{brand knowledge base (BKB)} and a \emph{detector backbone} utilizing this BKB.

A BKB \cite{phishpedia, phishintention, dynaphish} stores brand-related knowledge, taking the form of a list of $N$ brands: $b_1, \cdots, b_N$. For each brand $b$, we store its name (\emph{b.name}), its logo images (\emph{b.logos}), and its legitimate domains (\emph{b.domains}). In our work, to facilitate the detection of textual brand intention, we further add its list of textual aliases, i.e., a list of common alternate names used to refer to the brand (\emph{b.aliases}), resulting in an \emph{augmented BKB}. Formally, given a brand $b$, the augmented BKB is:
\begin{align*}
    \mathcal{B} = \{ (
    \textit{b.name},
    \textit{b.logos},
    \textit{b.domains},
    \textit{b.aliases} ) \}_{b \in \{ b_1, \cdots, b_N \} } 
\end{align*}
For example, for the brand \texttt{PayPal}, this may contain:   (\texttt{PayPal}, (\texttt{logo1}, \texttt{logo2}), (\texttt{www.paypal.com}), (\texttt{PYPL})), where \texttt{logo1} and \texttt{logo2} are two PayPal logo images.

Next, given a webpage $w$ and a BKB $\mathcal{B}$, a \emph{detector backbone} $g_\mathcal{B}(w)$ outputs either the brand intention that $w$ is predicted to have, or `null' to indicate no predicted brand intention: $g_\mathcal{B}(w) \in \mathcal{B} \ \cup \ \{ \varnothing \}$.

Finally, an RBPD, denoted $f_{\mathcal{B}}(w)$, classifies a webpage $w$ as phishing or benign. $f_{\mathcal{B}}$ classifies $w$ as phishing if its detector backbone $g_\mathcal{B}$ detects that $w$ presents a brand intention $b'\in\mathcal{B}$ but $w$'s domain is inconsistent with any of the legitimate domains of brand $b'$ recorded in $\mathcal{B}$ (i.e., $\mathit{w.domain}\notin \mathit{b'.domains}$). Otherwise, $w$ is classified as benign. Formally:
\begin{align*}
    f_{\mathcal{B}}(w) = 
\begin{cases} 
\text{Phishing} & \text{if } g_{\mathcal{B}}(w) = b' \text{ and } 
 b' \neq \varnothing \text{ and } \\ & \ \ \ w.\textit{domain} \notin b'.\textit{domains} \\ 
\text{Benign} & \text{otherwise}
\end{cases}
\end{align*}

\noindent\textbf{Evasion Attacks.}\ \ 
The attackers may attempt to bypass $f_{\mathcal{B}}$ via the following methods:

\noindent \emph{\textbf{T1: Phishing with Logo Variants.}} To circumvent the online knowledge expansion approach for $\mathcal{B}$ (e.g., DynaPhish\cite{dynaphish}), attackers can use other legitimate logo variants of $b$ instead of the ones displayed on $b$'s official webpages. 

\noindent \emph{\textbf{T2: Phishing with Text Brands.}} Instead of using a logo to present its brand intention, the attacker can rely on text \textit{w.html} to show its brand intention, making image-based phishing detectors completely fail.

\noindent \textcolor{revision_color}{\emph{\textbf{T3: HTML-oriented Attacks.}} Phishing attackers may employ evasion techniques on HTML, such as typosquatting, prompt injection, and text-to-image attacks to hinder effective information extraction by text-based methods.}

We address T1 by constructing a large-scale multimodal BKB with rich logo information, i.e., multiple logos per brand. We address T2 by developing an LLM-based approach to extract text brands from HTML. For T3, we show in \hyperref[sec:rq3]{Section \ref{sec:rq3}} that our LLM-based approach is generally robust to different types of adversarial noises in HTML.  
\section{KnowPhish Construction}
\label{sec:knowphish_construction}

In this section, we introduce how to construct \textit{KnowPhish}, a large-scale multimodal BKB. We start by conducting empirical analysis on phishing feeds from different \textcolor{revision_color}{sources and} periods to find prospective indicators to search for potential phishing targets proactively. With our empirical findings, we then describe how to automatically construct KnowPhish using a publicly accessible multimodal knowledge graph.

\subsection{Empirical Motivation}
\label{sec:empirical_motivation}
Our empirical analysis seeks to address the following questions: 

\textsc{Question 1}. \textit{\textcolor{revision_color}{Do phishing targets differ across different phishing feeds?}}

\textsc{Question 2}. \textit{What are the enduring characteristics shared by phishing feeds across different \textcolor{revision_color}{sources and periods}}?

\subsubsection{Data}
To achieve this objective, we conducted a study using two distinct phishing datasets. The first dataset, $D_1$, includes phishing webpages that were collected by \cite{phishpedia} \textcolor{revision_color}{from OpenPhish\cite{openphish}} three years ago. It encompasses a total of 29,496 phishing instances targeting 283 different brands. The second dataset, $D_2$, comprises phishing samples obtained from APWG\cite{apwg} at the end of 2022. This dataset contains a total of 5,167 phishing examples targeting 391 unique brands.

Furthermore, we manually categorized each of these phishing targets into one of the following distinct industries, namely: 1) financial, 2) online service, 3) telecommunication, 4) e-commerce, 5) social media, 6) postal service, 7) government, 8) web portal, 9) video game, and 10) gambling. For the brands that cannot be classified into any of the ten categories, we categorize them as 11) other businesses. For instance, Bank of America is categorized into the financial category, while KFC is classified as other businesses.

\subsubsection{Analysis}
\label{sec:empirical_analysis}
We conduct a thorough examination of the disparities in phishing targets and the distribution of their respective industries across the two phishing datasets. Our observations are as follows.

\begin{figure}[!t]
    \centering
    \includegraphics[scale=0.60]{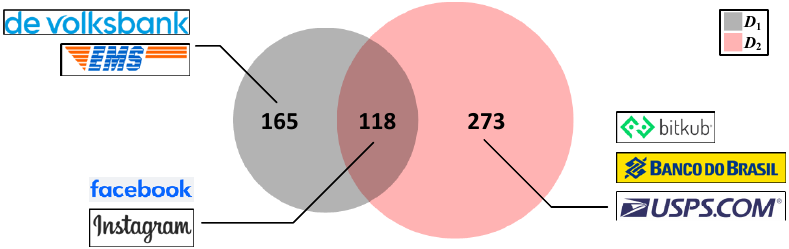}
    \caption{Phishing targets change substantially. The Venn diagram shows the disparities in the phishing targets from the two phishing datasets, with a few phishing target examples provided for illustration.}
    \label{fig:phishing_target_overlap}
\end{figure}
\begin{figure}[!t]
    \centering
    \includegraphics[scale=0.6]{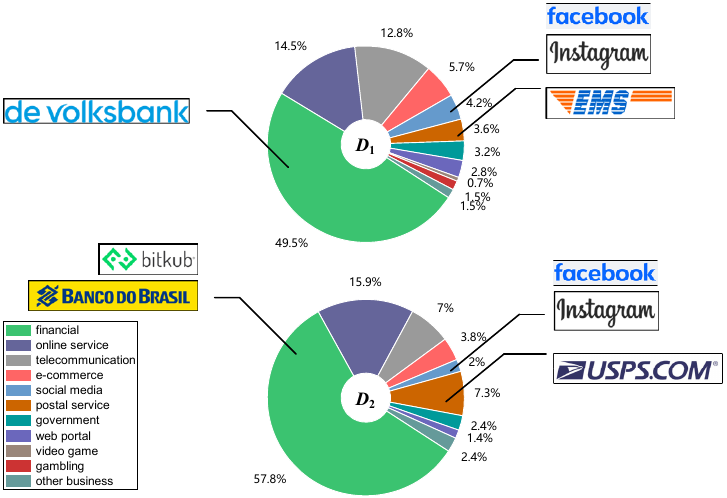}
    \caption{Industries of phishing target brands are relatively consistent. The chart shows the distribution of industries of the phishing targets from the two datasets.}
    \label{fig:phishing_target_industries}
\end{figure}

{\color{revision_color}{
\textsc{Observation 1.} \textit{Phishing targets may exhibit variation depending on the datasets involved.}


In particular, in \hyperref[fig:phishing_target_overlap]{Figure \ref{fig:phishing_target_overlap}} we observe significant phishing target disparities between $D_1$ and $D_2$, with only 118 of the 391 brands in $D_2$ being present in $D_1$. This can be attributed to many potential factors, including temporal shifts and data collection methodologies. On the temporal shifts side, emerging phishing targets (e.g., Bitkub) may supplant existing ones over time\cite{dynaphish}. On the data collection methodology side, OpenPhish employs proprietary detectors for phishing URL collection, whereas APWG relies on human reports. Such divergent approaches to data gathering directly impact the composition of phishing targets. These factors highlight the non-trivial challenge of manually managing the dynamic nature of phishing targets.

\textsc{Observation 2.} \textit{Despite the dynamic nature of phishing targets, the industries of those phishing targets remain mostly consistent.}
}}

However, an intriguing revelation from \hyperref[fig:phishing_target_industries]{Figure \ref{fig:phishing_target_industries}} is that despite the shifting landscape of phishing targets, the industries targeted by these phishing attacks have remained mostly consistent. For example, De Volksbank (in $D_1$), Bitkub, and Banco Do Brazil (in $D_2$) are all banking organizations, even though they \textcolor{revision_color}{are phishing targets from different sources}. Almost all the phishing targets in $D_2$ can still be categorized in the ten industries. The majority of these targets continue to belong to the same industries that have historically been the focus for phishing attacks, such as financial, telecommunication, and postal service, among others, while only a few brands such as KFC, Hydroquebec cannot be covered by the ten industries. Our supplementary materials\cite{knowphish_github} provide more examples of the phishing targets of different industries.

It is worth noting that this observation echoes the six Principles of Influence \cite{principles_of_influence} upon which social engineering relies, namely the significance of authority in achieving successful persuasion. In the context of social engineering, threat actors, including phishing attackers, tend to focus their efforts on authorized and higher-value entities to maximize their gains from unlawful acquisition of sensitive information, rather than impersonate less-reputable and lower-value firms.

\subsubsection{Connection to Wikidata Knowledge Graph} 
The consistent targeting of specific industries by phishing attacks can, in turn, serve as a valuable foundation for building a BKB predictively. The brand-industry relation can be regarded as a fact triplet stored within knowledge graphs. In this work, we use Wikidata \cite{wikidata}, the largest publicly-accessible knowledge base\cite{wikidata_largest_1, wikidata_largest_2} to explore the connection between phishing targets and the knowledge graph. Our focus lies on examining the \texttt{instance\_of} relationship within Wikidata, as we empirically find that it provides the most comprehensive information about the category to which an entity belongs. For example, the \textcolor{revision_color}{fact that} brand Bank of America belongs to the category bank can be represented as $(\text{bank\_of\_america}, \texttt{instance\_of}, \text{bank})$.  Formally, we use $(b, \texttt{instance\_of}, c)\in\mathcal{G}$ to represent that brand $b$ belongs to category $c$ if there is such a fact in the knowledge graph $\mathcal{G}$.  

\begin{figure}[!t]
    \centering
    \includegraphics[width=1\linewidth]{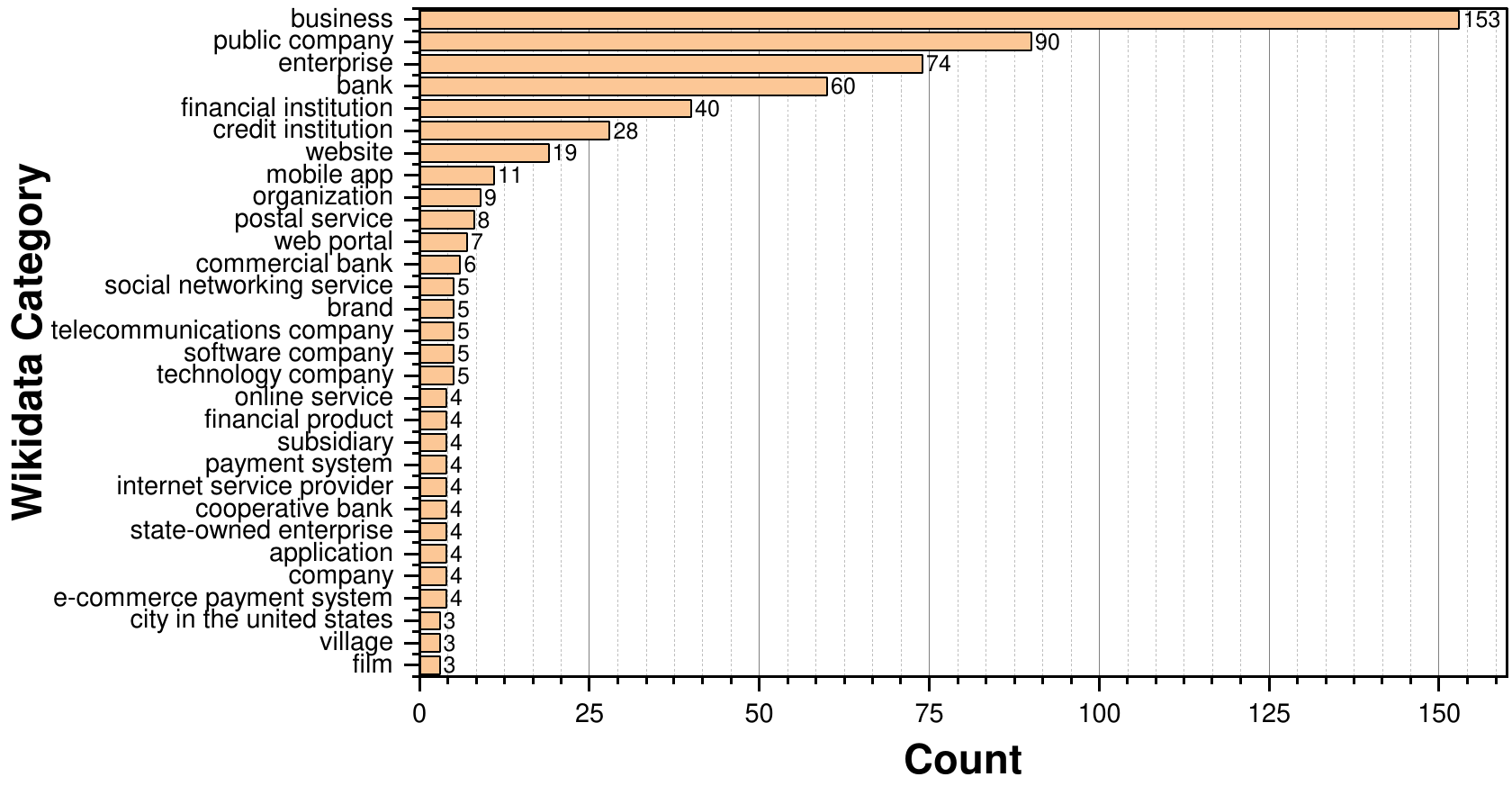}
    \caption{Distribution of the top 30 Wikidata categories of the phishing targets in $D_2$.}
    \label{fig:instance_of}
\end{figure}

To gain a deeper understanding of which Wikidata categories those phishing targets belong to, we perform searches within the $D_2$ dataset. Specifically, for each phishing target $b$, we search for the categories associated with it, denoted as $\mathcal{C}(b)=\{c|(b, \texttt{instance\_of}, c)\in\mathcal{G}\}$.  This process yields a collection of categories, $\mathcal{C}$, comprising the categories for all phishing targets in the APWG phishing dataset.  In \hyperref[fig:instance_of]{Figure \ref{fig:instance_of}}, we present a visualization of the 30 most frequently occurring categories within $\mathcal{C}$. Our observation reveals that some categories represent specific industries, such as bank, postal service, and internet service provider, while others, like business, public company, and enterprise, convey more general semantics.

Based on this observation, we curate two lists of categories: 1) \emph{Narrow Categories} $\mathcal{C}_n$ such as `bank' representing narrower industry segments, and 2) \emph{General Categories} $\mathcal{C}_g$ such as `business', which will be used in constructing our brand knowledge base KnowPhish. The detailed lists can be found in our supplementary materials\cite{knowphish_github}.  

\subsection{Approach Overview}
\begin{figure*}[t]
    \centering
    \includegraphics[scale=0.65]{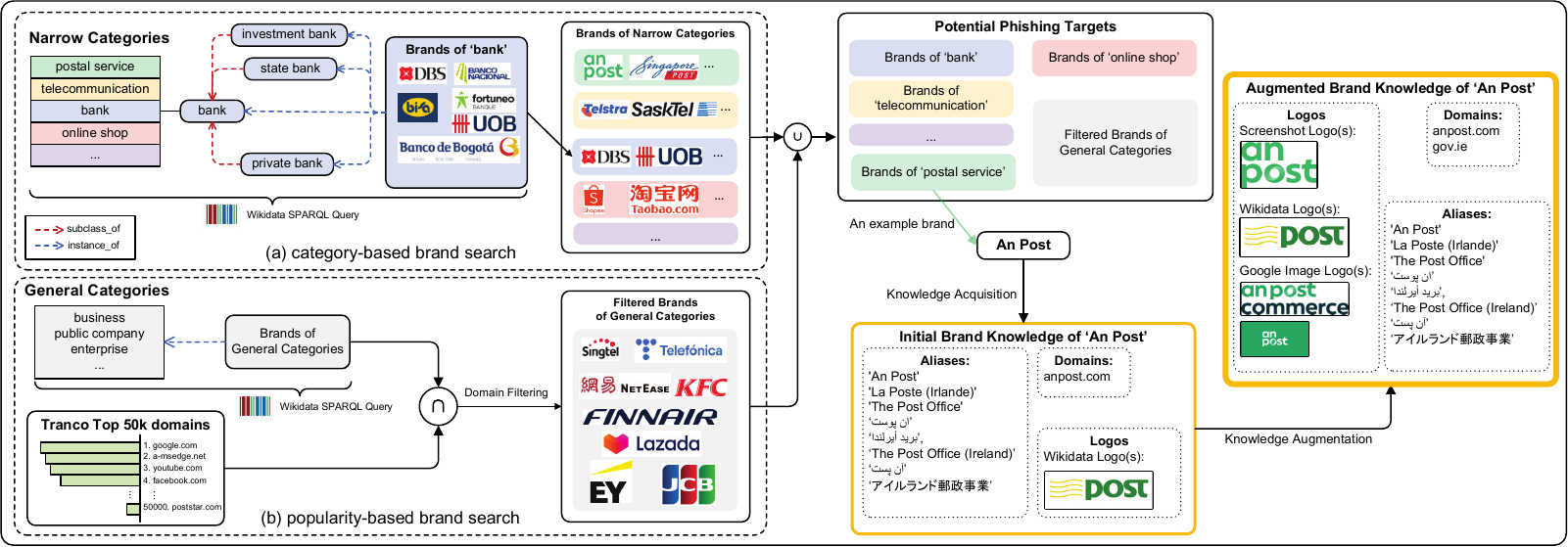}
    \caption{An overview of our automated pipeline for constructing our large-scale multimodal BKB, KnowPhish. We first collect (a) all brands from certain high-value industries, and (b) only popular brands from general categories. Then, the knowledge acquisition and augmentation steps collect logos, domains, and aliases for these brands.}
    \label{fig:knowphish_construction}
\end{figure*}
Building upon the empirical insights in \hyperref[sec:empirical_motivation]{Section \ref{sec:empirical_motivation}}, we introduce \textit{KnowPhish}, a large-scale multimodal BKB prioritizing both comprehensive coverage of potential phishing targets and detailed brand knowledge. As \hyperref[fig:knowphish_construction]{Figure \ref{fig:knowphish_construction}} shows, KnowPhish is constructed using an automatic pipeline, which starts from \textbf{1) Brand Search} based on industries via Wikidata knowledge graph $\mathcal{G}$ and proceeds with \textbf{2) Knowledge Acquisition and Augmentation} to obtain relevant brand knowledge including logos, aliases, and domains. The complete construction algorithm is illustrated in our supplementary materials\cite{knowphish_github}.

\subsection{Brand Search}
Identifying a wide range of potential phishing targets is crucial for the construction of KnowPhish. If a phishing target is absent from the BKB, the RBPD may not be able to identify its corresponding phishing webpages, resulting in false negatives. Our brand search module is designed to prioritize brands that are higher-value phishing targets and consists of two concurrent components:
\begin{itemize}
    \item[(a)] \textbf{Category-based Brand Search} identifies brands operating within specific industries, i.e., Narrow Categories $\mathcal{C}_n$, to find brands $\mathcal{B}_n$.
    \item[(b)] \textbf{Popularity-based Brand Search} considers a broader set of General Categories $\mathcal{C}_g$ and ranks their brands by popularity to generate a brand list $\mathcal{B}_g$.
\end{itemize}
By combining (a) and (b), we obtain a more comprehensive list of potential phishing targets, denoted as $\mathcal{B}=\mathcal{B}_n \cup \mathcal{B}_g$.

\subsubsection{Category-based Brand Search}
Category-based brand search is motivated by our empirical observation that phishing attackers often choose to impersonate brands within high-value industries. Thus, we collect all brands associated with Narrow Categories $\mathcal{C}_n$. For instance, when focusing on the `bank' category, we aim to find all the banks listed in Wikidata. 

Hence, for each Narrow Category $c_n\in\mathcal{C}_n$, we search for the corresponding brands belonging to $c_n$ and its subcategories $\mathcal{C}_n' = \{c | (c, \texttt{subclass\_of}, c_n)\in\mathcal{G}\}$, where \texttt{subclass\_of} indicates a hierarchy relationship between two categories in the Wikidata graph $\mathcal{G}$. Formally, the brand list related to category $c_n$ is;
\begin{equation*}
    \mathcal{B}_n(c_n) = \{b|(b, \texttt{instance\_of}, c)\in\mathcal{G}, c\in\{c_n\} \cup \mathcal{C}_n'\}
\end{equation*}
We introduce sub-categories here because the brands within a sub-category of a Narrow Category are also potential targets for phishing attackers. For example, National Bank of Costa Rica is only categorized as `state bank', a sub-category of `bank', in Wikidata. Despite this, it unquestionably falls under the category of `bank', and is a potential phishing target. 

\subsubsection{Popularity-based Brand Search}
Although the industry category is a key indicator of potential phishing targets, such information may not be accurately maintained in Wikidata for every brand. For example, Singtel, a Singaporean telco company, should logically fall into the telecommunication category. However, the only categories it has in Wikidata are `business', `public company', and `enterprise'. Hence, depending solely on Narrow Categories $\mathcal{C}_n$ is insufficient.

To address this, we introduce popularity-based brand search that incorporates \emph{domain rank} for selecting brands. We use domain rank-based filtering for two reasons: 1) not all the brands within General Categories $\mathcal{C}_g$ are high-value entities, and 2) the more reputable a brand is, the more likely it is to be a phishing target. As a result, for each general category $c_g \in \mathcal{C}_g$, we generate its corresponding brand list with a domain ranking constraint:
\begin{align*}
 \begin{split}
     \mathcal{B}_g(c_g) = \{b | (b, \texttt{instance\_of}, c_g)\in\mathcal{G},
     r_{\mathcal{D}}(b.domains)\le\eta\}
 \end{split}
 \end{align*}
 where $\mathcal{D}$ is a popular domain ranking list, $r_{\mathcal{D}}(\cdot)$ uses $\mathcal{D}$ to compute the domain rank of the most popular domain in \textit{b.domains}, and $\eta$ is a domain ranking threshold. Here, we instantiate $\mathcal{D}$ with the Tranco domain ranking list \cite{tranco}. 

The brands obtained through the category-based brand search, denoted as $\mathcal{B}_n$, and popularity-based brand search, denoted as $\mathcal{B}_g$, are combined into our final list $\mathcal{B}=\mathcal{B}_n \cup \mathcal{B}_g$.

\subsection{Knowledge Acquisition and Augmentation}
\label{sec:knowledge_acquisition_and_augmentation}
RBPDs fundamentally depend on their brand knowledge to allow for accurate phishing detection. Next, we augment our collected brands with knowledge about the 1) \textit{logos}, 2) \textit{domains}, and 3) \textit{aliases} (or alternate names) associated with each brand. Note that aliases are not present in existing BKBs~\cite{phishpedia,phishintention,dynaphish}, but we introduce them to facilitate the detection of textual brand intention, as we describe in \hyperref[sec:knowphish_detector]{Section \ref{sec:knowphish_detector}}.

\subsubsection{Knowledge Acquisition}
Each brand $b \in \mathcal{B}$ we have collected is a Wikidata entity with rich property information. Therefore, we leverage this readily available data to establish initial brand knowledge. Specifically, for each $b \in \mathcal{B}$, we acquire initial brand knowledge from the Wikidata graph $\mathcal{G}$:
\begin{align*}
    & b.logos \leftarrow \{x | (b, \texttt{logo\_image}, x) \in \mathcal{G}\} \\
    & b.domains \leftarrow \{y.domain | (b, \texttt{official\_website}, y) \in \mathcal{G}\} \\
    & b.aliases \leftarrow \{z | (b, \texttt{label}, z) \in \mathcal{G}\}
\end{align*}
where \texttt{logo\_image}, \texttt{official\_website}, and \texttt{label}
are the property relations in $\mathcal{G}$ that indicate the logos, URL of the official website, and alternative names in different languages of an entity, respectively.  

\subsubsection{Knowledge Augmentation}
Information maintained in Wikidata may be incomplete, particularly for the logos and the domains. Brands may employ multiple legitimate logo variants and domain variants in their online presence. When a phishing page contains a logo variant not present in the knowledge base, the phishing detector may fail to identify its brand, leading to false negatives. Similarly, if our detector examines a benign webpage with a legitimate domain that is not documented in Wikidata, a false positive alarm may be raised. Thus, further augmentation to $\mathcal{B}$ on the logos and domains is required to alleviate such false positives and false negatives. 

\noindent\textbf{Logo Variants.}\ \ 
To capture logo variants, we employ two methods. The first involves accessing the associated domain(s) of the brand and capturing the logo displayed on that webpage by a well-trained webpage layout detector\cite{phishintention}, denoted $\textsf{DetectLogos}(b.domains)$. The second method utilizes Google Image Search\cite{google_images_search}. We initiate a search query by combining the brand name with the term `logo', then filter the results to include images with URLs matching the brand's domain(s). In this way, we expand our logo collection beyond the Wikidata logo images:
\begin{equation*}
\begin{split}
    b.logos \leftarrow b.logos \ \cup \  \textsf{DetectLogos}(b.domains)\\
    \ \cup \  \textsf{GoogleImageLogos}(b.name + \text{`logo'})
\end{split}
\end{equation*}

\noindent\textbf{Domain Variants.}\ \ 
To acquire additional domain variants, we utilize the Tranco domain ranking list $\mathcal{D}$ and the Whois service\cite{whois}. Concretely, we run the Whois lookup on all the domains in KnowPhish and $\mathcal{D}$ to gather the Whois information for each of their domains. Then for each brand $b$, we expand the list of its legitimate domains by incorporating domains in $\mathcal{D}$ that share identical organization details with the original $b.domains$:
\begin{equation*}
\begin{split}
    b.domains \leftarrow b.domains \cup \{d | h_{whois}(d).org =\\
    h_{whois}(b.domains).org,\ d\in\mathcal{D}\}
\end{split}
\end{equation*}
Here, $h_{whois}(d)$ refers to the Whois information for domain $d$. Note that the organization entry in the Whois information specifies the owner of a domain. Therefore, domains within $\mathcal{D}$ owned by the same entity can effectively complement our list of domain variants.

\noindent\textbf{Domain Propagation.}\ \ 
We further propose a method for propagating domain information among brand pairs that share subsidiary relationships, since the legitimate website of a brand may also display the logos of its subsidiary (or vice versa). For instance, `facebook.com' can be seen as a domain variant for Meta. When visiting `facebook.com', it is reasonable for a Meta logo to be present; but if we were not aware of this domain variant, we would classify it as having the brand intention of Meta, and thus a phishing attack, resulting in a false positive.

To address this problem, we use the subsidiary relationship under the \texttt{owned\_by} and \texttt{parent\_organization} property relations in $\mathcal{G}$. For each $b \in \mathcal{B}$, its `propagated domains' is defined as all domains in its 1-hop neighborhood over the graph of these relations; that is:
\begin{align*}
     & b.domains \cup \{b'.domains| b'\in\mathcal{N}(b), b'\in\mathcal{B}\}, \text{ where } \\
    \mathcal{N}(b) &= \{b' | (b, \texttt{owned\_by}, b')\in\mathcal{G}\vee(b', \texttt{owned\_by}, b)\in\mathcal{G} \\
    & \vee(b, \texttt{parent\_organization}, b')\in\mathcal{G} \\
    & \vee(b', \texttt{parent\_organization}, b)\in\mathcal{G}\}
\end{align*}
represents the domains from a collection of brands that share subsidiary relationships with the brand $b$. At the end of domain propagation, we replace the original domains ($\textit{b.domains}$) with the propagated domains as above.

By now KnowPhish has been constructed completely and is ready to be equipped with image-based \cite{phishpedia, phishintention} and text-based phishing detectors (discussed in \hyperref[sec:knowphish_detector]{Section \ref{sec:knowphish_detector}}).

\noindent\textbf{Adapting to Evolving Phishing Targets.}
New brands are continuously emerging as potential phishing targets. Since KnowPhish is built in a fully automatic manner, one can simply handle such information obsolescence by regularly reconstructing KnowPhish to search for new potential phishing targets outside $\mathcal{B}$. Such regular updates allow the RBPD method to remain effective in countering attacks targeting these emerging brands.

{\color{revision_color}{
\noindent\textbf{Adversarial Injection}\ \ 
Despite Wikidata being well-maintained, there remains a risk of attackers inserting phishing URLs into its database. To mitigate this, we verify URLs against existing phishing blacklists (e.g., Google Safe Browsing\cite{google_safe_browsing}) before adding them to our BKB.
}}

\section{KnowPhish Detector}
\label{sec:knowphish_detector}

\begin{figure}[!t]
    \centering
    \includegraphics[scale=0.45]{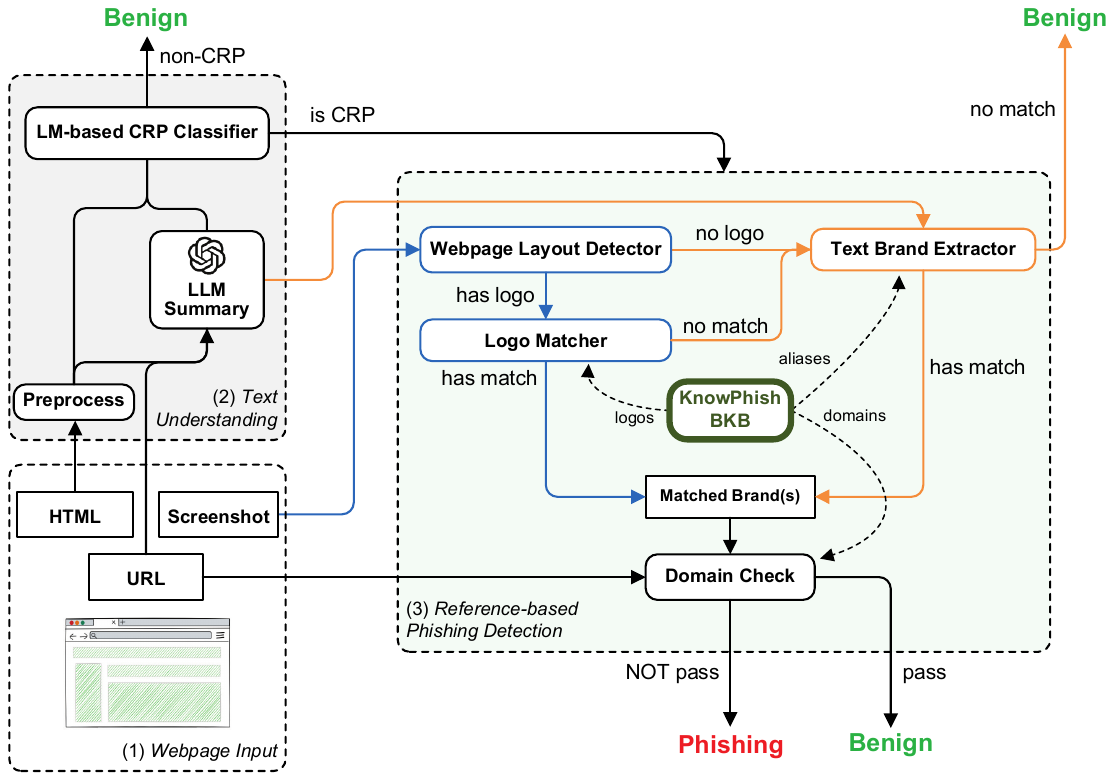}
    \caption{An overview of our phishing detector KPD.}
    \label{fig:knowphish_detector}
\end{figure}

Incorporating our multimodal BKB KnowPhish, we further propose \textit{KnowPhish Detector (KPD)}, a multimodal RBPD with multi-stage analysis. \hyperref[fig:knowphish_detector]{Figure \ref{fig:knowphish_detector}} offers an overview of KPD, and \hyperref[algo:knowphish_detector_algo]{Algorithm \ref{algo:knowphish_detector_algo}} further elaborates on its analysis steps. Specifically, for an input webpage $w$, KPD first leverages an LLM to generate a summary for $w$ using its HTML and URL. Then, the summary and the HTML are fed into a well-trained small language model to classify whether $w$ is a credential-requiring page (CRP). If $w$ is detected as a CRP, KPD will proceed to extract its brand intention from either the screenshot or the LLM summary. The extracted brand intention of $w$ is then used to retrieve a list of legitimate domains, which are compared with the domain of $w$ to decide whether $w$ is phishing or not. Notably, KPD features novel text-based modules to analyze implicit CRP and extract brand intention from logo-less webpages, which will be discussed in \hyperref[sec:text_crp]{Section \ref{sec:text_crp}} and \hyperref[sec:text_brand]{Section \ref{sec:text_brand}}, respectively. 

\subsection{LLM-based Webpage Summary}
The LLM summary, as depicted in \hyperref[fig:knowphish_detector]{Figure \ref{fig:knowphish_detector}}, acts as crucial information for various subsequent tasks. To generate this summary, we first process the HTML by removing extraneous elements like JavaScript and CSS blocks. Subsequently, this processed HTML, along with the URL, will become the input of an LLM. We design a prompt template (details provided in \cite{knowphish_github}) with three in-context examples to understand the webpage from various aspects, such as what the brand intention of the webpage is, whether the webpage is a CRP, which elements in the HTML makes it a CRP, and the overall rationale for its CRP prediction. The generated summary will serve as an auxiliary text attribute of the webpage for the following  CRP classification and text brand extraction tasks.

\begin{algorithm}[t!]\footnotesize
\renewcommand{\arraystretch}{.7}
    \caption{KnowPhish Detector (KPD)}
    \label{algo:knowphish_detector_algo}
    \SetKwInOut{Input}{Input}
    \SetKwInOut{Output}{Output}
    \SetKwInOut{Notation}{Notations}
    \SetAlgoLined
    
    \newcommand\mycommentfont[1]{\texttt{\textcolor{blue}{#1}}}
    \SetCommentSty{mycommentfont}
    
    \Input{Webpage $w =(\textit{w.html}, \textit{w.screenshot}, \textit{w.domain})$, LLM-based webpage summarizer \textsf{LLMSummary}, text-based CRP classifier \textsf{CRPClassifier}, logo brand extractor \textsf{LogoBrand}, text brand extractor from LLM summary \textsf{TextBrand}, alias map $\textsf{AliasMap}: a \rightarrow \{b|b \in \mathcal{B} \wedge a \in b.aliases\}$ mapping a text string $a$ to the set of brands in KnowPhish that has $a$ as its alias.}
    \Output{Whether $w$ is phishing or benign}

    
    $s \leftarrow \textsf{LLMSummary}(\textit{w.html}, \textit{w.domain})$\;
    \tcp{\footnotesize{If no CRP detected, treat as benign}}
    \If{$\textsf{CRPClassifier}(s, \textit{w.html})$ is False}{
        \Return Benign;
    }
    \tcp{\footnotesize{Extract brand first visually, then textually by parsing LLM summary}}
    $b_v \leftarrow \textsf{LogoBrand}(\textit{w.screenshot})$; \\
    \If{$|b_v|=0$}{
        $b_t \leftarrow \textsf{AliasMap}(\textsf{TextBrand}(\textit{s}))$
    }
    
    \tcp{\footnotesize{If no brand was matched, then treat as benign}}
    \If{$|b_v|=0$ and $|b_t|=0$}{
        \Return Benign;
    }

    \tcp{\footnotesize{If current page domain is a legitimate domain of any extracted brand, it is benign}}
    \For{$b \in b_v \ \cup \ b_t $}{
        \If{$\textit{w.domain} \in \textit{b.domains}$}{
            \Return Benign;
        }
    }
    \Return Phishing;
\end{algorithm}

\subsection{Text-based CRP Classification}
\label{sec:text_crp}

\begin{figure}[!t]
    \begin{minipage}[b]{0.495\linewidth}
    \centering
    \flushbottom
    \subfigure[Explicit CRP]{
        \includegraphics[scale=0.095]{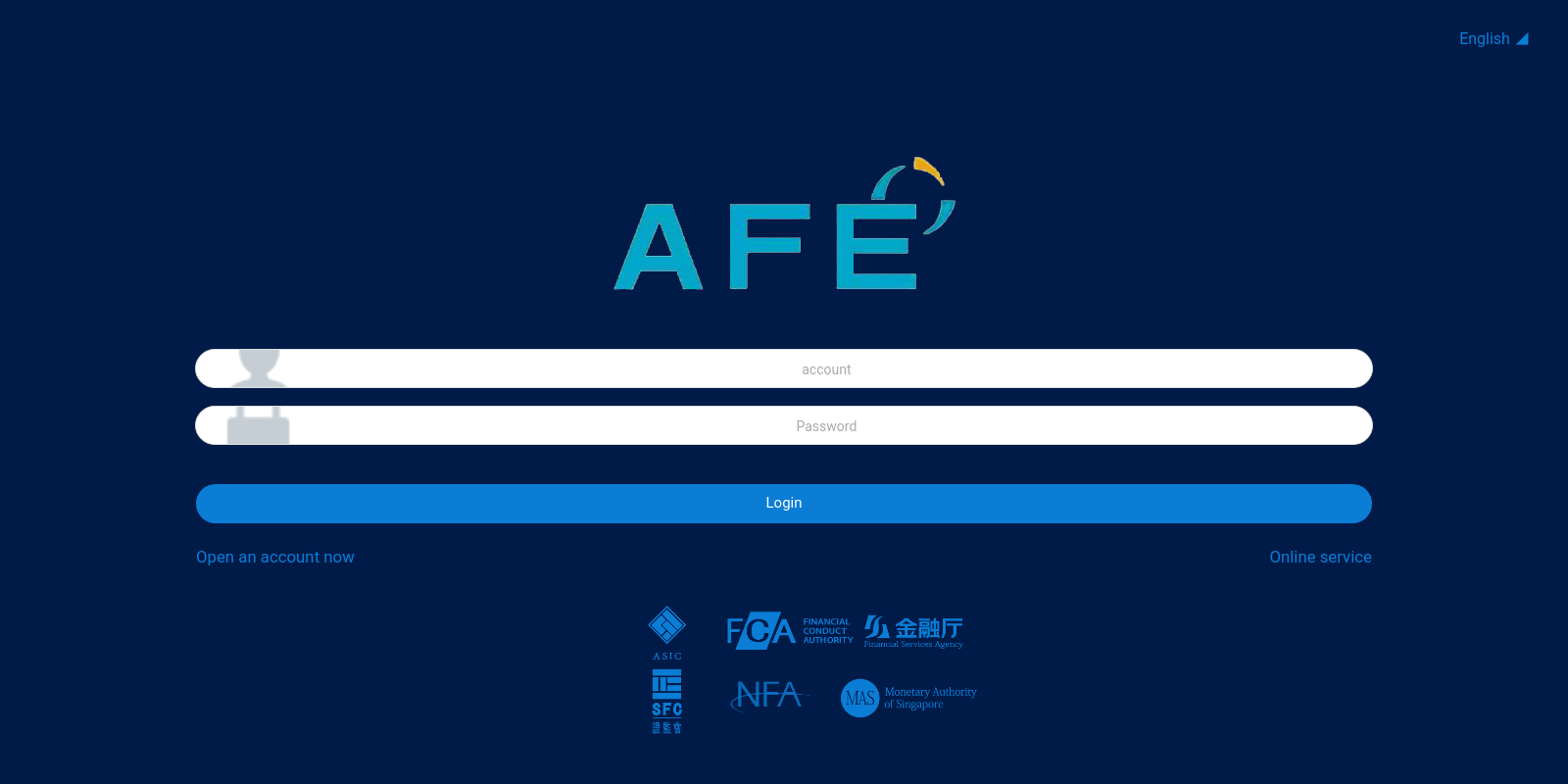}
    }
    \end{minipage}
    \begin{minipage}[b]{0.495\linewidth}
    \centering
    \flushbottom
    \subfigure[Implicit CRP]{
        \includegraphics[scale=0.095]{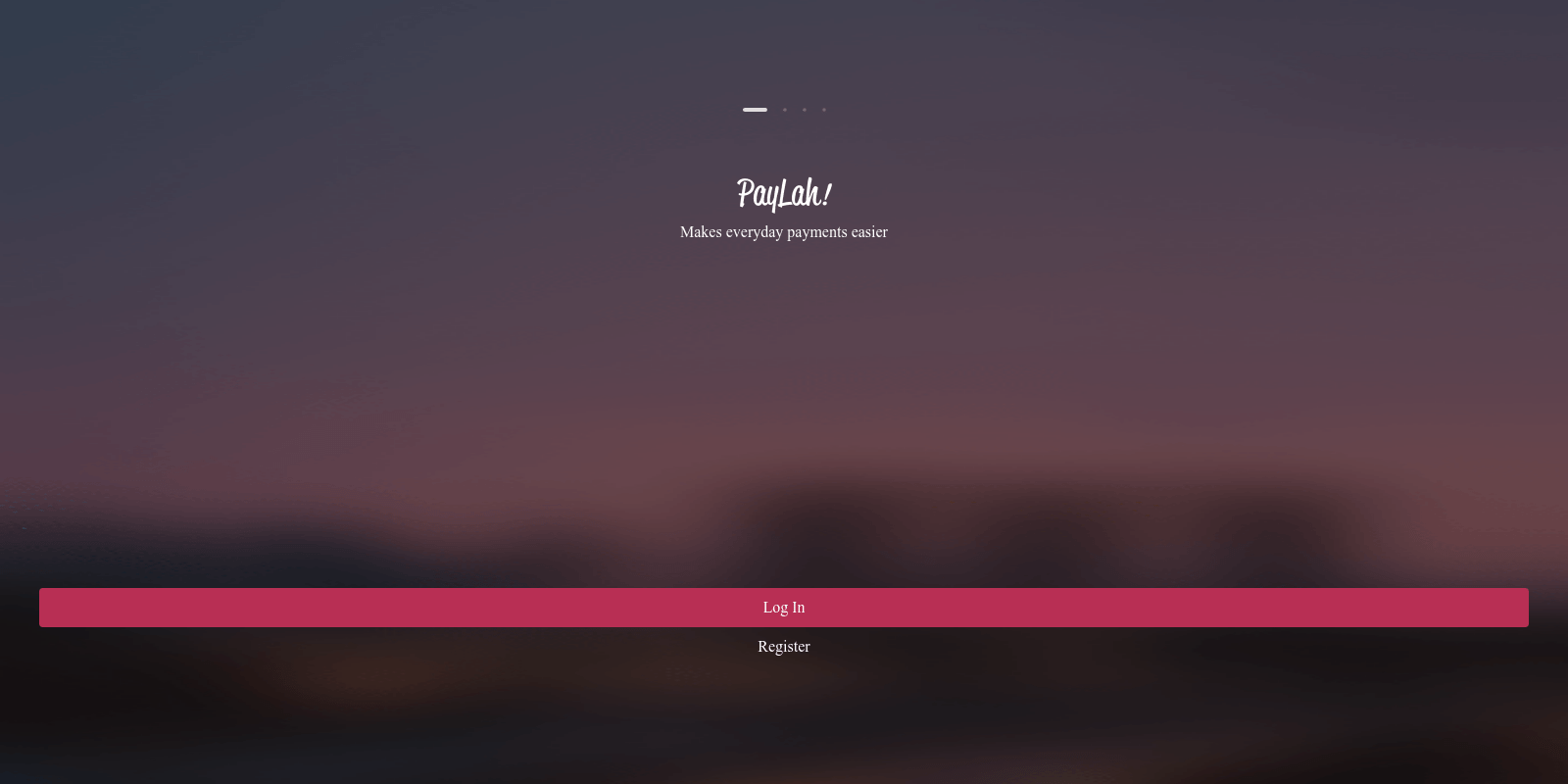}
    }
    \end{minipage}
    \caption{Comparison between explicit and implicit CRP.}
    \label{fig:crp_comparison}
\end{figure}

As discussed in \hyperref[sec:formalization]{Section \ref{sec:formalization}}, phishing webpages always convey credential-requiring intentions. While a prior study has already developed an image-based CRP classifier\cite{phishintention}, we empirically find that it leads to a significant number of false negatives during deployment by incorrectly classifying phishing webpages into non-CRP, in which most of these false negatives are implicit CRP. \hyperref[fig:crp_comparison]{Figure \ref{fig:crp_comparison}} provides a comparison between explicit and implicit CRP, where explicit CRP directly shows credential-taking elements (e.g., the input fields of username and passwords), whereas implicit CRP only presents the elements (e.g., a login button) to redirect to a potential explicit CRP. While the credential-requiring intention is evident to page visitors, the CRP classifier from \cite{phishintention} cannot identify it because it exclusively focuses on identifying explicit CRPs.

In response to this limitation, we propose using a small \textcolor{revision_color}{multilingual} LM (XLM-RoBERTa\cite{xlmroberta}) to identify credential-requiring intention from text. Specifically, the small LM takes both the processed HTML and the LLM summary as input and outputs a binary prediction of whether the webpage is a CRP or not. This design of integrating the original text and its LLM summary as input to a smaller LM aims to benefit from the general-purpose reasoning and instruction following ability of an LLM, along with the trainability of a small LM~\cite{tape}. Consequently, our text-based classifier can better comprehend the credential-requiring intention from webpage text, facilitating the detection of both explicit and implicit CRP. Webpages classified as non-CRP are regarded as benign; for those classified as CRP, we proceed to the brand extractor step.

\subsection{Brand Extractor}
\label{sec:text_brand}
Next, we aim to extract the brand intention of the webpage. Recall that our approach integrates with an existing RBPD which identifies brands visually through logos, which we call a \emph{logo brand extractor (LBE)}. Specifically, existing LBEs~\cite{phishpedia, phishintention} consists of a \emph{webpage layout detector} to locate a logo from the screenshot, and a \emph{logo matcher} to match that logo to a brand in the BKB. We find that existing RBPDs encounter limitations in identifying brand intention when 1) the logo displayed on the webpage differs from the logos stored in the BKB, or 2) the logo cannot even be detected.

To cope with this problem, we introduce a \emph{text brand extractor (TBE)}, which acts as an extra component when the LBE fails to identify a brand from the screenshot. TBE directly extracts the text brand by parsing the LLM summary which already contains a brand intention prediction. The predicted text brand undergoes an exact matching process with all the aliases in KnowPhish. The brand associated with the matched alias becomes the identified brand during the brand identification step. In the event of multiple aliases matching, all corresponding brands become identified.

In situations where the LBE fails to detect a logo on the webpage or cannot find a matching logo, the TBE is activated to extract brand intention, leading to higher recall. 
\textcolor{revision_color}{
TBE is similar to the counterfactual analysis module \cite{dynaphish} in its ability to detect logo-less pages. However, the counterfactual analysis focuses on interacting with the webpage, e.g. by verifying credentials or redirection, while TBE focuses on detecting brand intention through the textual information on the webpage.
}

\subsection{Domain Check} Once both the credential-requiring intention and the brand intention are confirmed, the final step is to perform a domain check. We retrieve all the legitimate domains of the matched brand(s) from KnowPhish and compare them with the domain of the input webpage. If the input domain is inconsistent with all the legitimate domains we retrieve, the webpage will be classified as phishing; otherwise, it is predicted as benign. Note that webpages classified as non-CRP or having no brand intention are also deemed benign.
\section{Experiments}

We conduct experiments to answer the following research questions:
\begin{itemize}
    \item \textbf{[RQ1] Effectiveness and Efficiency}: Can KnowPhish and KPD effectively improve the phishing detection performance of existing phishing detectors?
    \item \textbf{[RQ2] Field Study}: How effective are KnowPhish and KPD when deployed in real-world scenarios?
    \item \textbf{[RQ3] Adversarial Robustness}: How robust is the text-based phishing detector against adversarial noise in HTML texts?
    \item \textbf{[RQ4] Ablation Studies}: How does each component of KPD contribute to its overall performance?
\end{itemize}

\subsection{Datasets}
We utilize two datasets for our main phishing detection experiments. 1) \texttt{TR-OP}: A manually labeled and balanced dataset where the benign samples are randomly collected from Tranco\cite{tranco} \textcolor{revision_color}{top 50k domains} and the phishing samples are obtained from OpenPhish\cite{openphish}. The phishing samples were crawled and validated within 6 months from July to December 2023, covering 440 unique phishing targets. Note that the phishing samples here are different from $D_1$ and $D_2$ discussed in \hyperref[sec:empirical_motivation]{Section \ref{sec:empirical_motivation}}. 2) \texttt{SG-SCAN}: An unlabelled dataset with samples from Singapore's local webpage traffic. We randomly sample 10k webpages dating from mid-August 2023 to mid-January 2024. It is used to evaluate the phishing detection approaches in the local context. \hyperref[tab:dataset]{Table \ref{tab:dataset}} offers an overview of both datasets.

\begin{table}[htbp]\footnotesize
    \centering
    \renewcommand{\arraystretch}{0.9}
    \begin{tabular}{ccccc}
    \toprule
        \textbf{Dataset} & \textbf{\#Samples} & \textbf{\#Benign} & \textbf{\#Phishing} & \textbf{Used in} \\
    \midrule
        \texttt{TR-OP} & 10k & 5000& 5000 & RQ1,3, and 4\\
        \texttt{SG-SCAN} & 10k & Unknown & Unknown & RQ2 and 4\\
    \bottomrule
    \end{tabular}
    \caption{Statistical overview of the main datasets.}
    \label{tab:dataset}
\end{table}

In addition, we also manually extracted and labeled 2555 samples to train an XLM-RoBERTa \cite{xlmroberta}, our text-based CRP classifier. This dataset contains 1094 phishing samples from $D_2$ and 1461 benign samples from Alexa Ranking \cite{alexa}. The 1094 phishing samples are all CRP. Among the 1461 benign samples, 1297 are non-CRP, while the remaining 164 are CRP. After combining these samples, they are divided into 0.8/0.1/0.1 train/valid/test splits.

\subsection{Baselines}
\label{sec:baselines}
We select two state-of-the-art approaches, Phishpedia\cite{phishpedia} and PhishIntention\cite{phishintention}, together with our proposed KPD as the RBPD backbones. Both Phishpedia and PhishIntention can be either equipped with their original reference list (containing 277 brands), DynaPhish\cite{dynaphish}, or our proposed KnowPhish, as the BKB used for phishing detection. \textcolor{revision_color}{Due to the requirement of alias information, KPD will be only equipped with KnowPhish or an extended version of DynaPhish. In this extended DynaPhish, the extracted brand name will be used as the only alias of each new brand.} For fair comparison, both KnowPhish and DynaPhish will construct their knowledge from an empty BKB, since one can always improve the performance of both knowledge expansion approaches by manually adding well-inspected brand knowledge.

We assume a static environment in all our experiments (i.e., the only data available on a webpage is its URL, screenshot, and HTML). In this case, the dynamic analysis module in PhishIntention and the webpage interaction module in DynaPhish will be disabled. Further details on the implementation can be found in \hyperref[app:implementation]{Appendix \ref{app:implementation}}.

\subsection{RQ1: Effectiveness and Efficiency}
\label{sec:rq1}


We evaluate the effectiveness of different RBPDs via accuracy, F1 score, precision, recall, number of brands detected, and efficiency based on the average running time per sample. Specifically, the number of brands detected is useful to understand how many unique brands each RBPD can identify from the phishing webpages, since identifying the target of a webpage is a crucial task for RBPDs.

\begin{figure}[!t]
    \centering
    \includegraphics[scale=0.25]{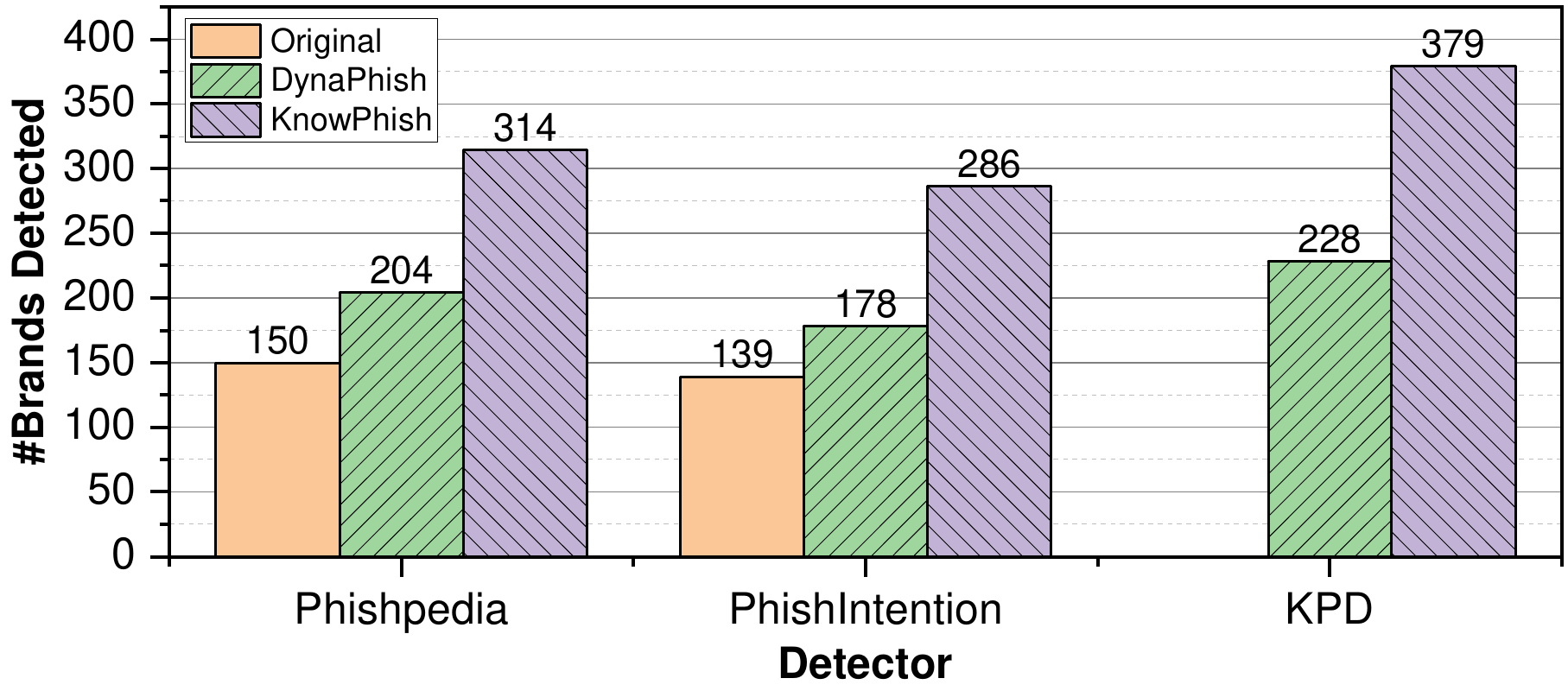}
    \caption{Comparison of the number of phishing targets detected by different RBPDs on \texttt{TR-OP} dataset.}
    \label{fig:phishing_targets_detected}
\end{figure}

\hyperref[tab:effectiveness_and_efficiency]{Table \ref{tab:effectiveness_and_efficiency}} shows the phishing detection performance of the three RBPDs with different BKBs. We observed the following key advantages of KnowPhish and KPD:
\begin{itemize}
    \item KnowPhish substantially boosts the F1 score of Phishpedia by 25\% and PhishIntention by 20\%, and also increases their recall by 32\% and 22\%, with only marginal impacts on precision, compared to other BKB baselines. The primary factor contributing to the superior performance of KnowPhish over DynaPhish is that it already encompasses most phishing targets and their logo variants. \textcolor{revision_color}{In contrast, DynaPhish suffers from the logo-matching constraint required to build brand knowledge, causing many false negatives as illustrated in \hyperref[fig:comparison]{Figure \ref{fig:comparison}}}. 
    
    \item KPD\textcolor{revision_color}{+KnowPhish} provides the highest F1 score of 92.05\%, and recall of 86.90\%, substantially outperforming other approaches. KPD benefits from the rich alias information from KnowPhish, allowing it to detect logo-less phishing pages through analysis of HTML and URL. Consequently, KPD identifies more phishing targets than DynaPhish, as shown in \hyperref[fig:phishing_targets_detected]{Figure \ref{fig:phishing_targets_detected}}. 
    
    \item KnowPhish achieves better runtime efficiency than DynaPhish by decoupling the BKB construction from phishing detectors. Unlike DynaPhish which requires crawling additional webpages during deployment, KnowPhish identifies potential phishing targets locally, leading to about 50 times lower running time when integrated with Phishpedia and PhishIntention. Even when equipped with KPD with additional LLM query overhead, \textcolor{revision_color}{KnowPhish remains 6 times faster than DynaPhish}. 
\end{itemize}
\begin{table}[htbp]\scriptsize
    \centering
    \renewcommand{\arraystretch}{0.9}
    \begin{tabular}{p{1.1cm}p{1.0cm}p{0.6cm}<{\centering}p{0.6cm}<{\centering}p{0.85cm}<{\centering}p{0.6cm}<{\centering}p{0.6cm}<{\centering}}
    \toprule
        \textbf{Detector} & \textbf{BKB} & \textbf{ACC}\textcolor{red}{$\uparrow$} & \textbf{F1}\textcolor{red}{$\uparrow$} & \textbf{Precision}\textcolor{red}{$\uparrow$} & \textbf{Recall}\textcolor{red}{$\uparrow$} & \textbf{Time}\textcolor{cyan}{$\downarrow$}\\
    \midrule    
                            & Original          & 69.91 & 57.17 & 99.16 & 40.16 & 0.25s     \\
         Phishpedia         & DynaPhish         & 66.40 & 52.52 & 89.50 & 37.16 & 10.92s    \\
                            & KnowPhish         & 85.79 & 83.67 & 98.27 & 72.80 & 0.22s     \\
    \midrule
                            & Original          & 66.62 & 49.96 & \textbf{99.76} & 33.32 &  0.28s \\
        PhishIntention      & DynaPhish         & 62.51 & 41.16 & 95.62 & 26.22 & 10.67s     \\ 
                            & KnowPhish         & 77.84 & 71.60 & 99.67 & 55.84 &  0.26s     \\
    \midrule
        \multirow{2}{*}{KPD} & \textcolor{revision_color}{DynaPhish}  &  \textcolor{revision_color}{76.10}  & \textcolor{revision_color}{69.71} & \textcolor{revision_color}{95.16} & \textcolor{revision_color}{55.00} & \textcolor{revision_color}{12.18s} \\
                            & KnowPhish         & \textbf{92.49} & \textbf{92.05} & 97.84  & \textbf{86.90} & 2.02s     \\
                            
    \bottomrule
    \end{tabular}
    \caption{Phishing detection performance of different RBPDs on \texttt{TR-OP} dataset. The metric `Time' indicates the average inference time per sample, while the remaining metrics are presented in percentages. \textcolor{red}{$\uparrow$} means higher is better while \textcolor{cyan}{$\downarrow$} refers to the opposite.}
    \label{tab:effectiveness_and_efficiency}
\end{table}

Overall, KnowPhish not only enhances brand coverage but also enables effective detection of logo-less phishing pages using KPD. This substantially improves the detection performance \hypertarget{target:efficiency1}{}\textcolor{revision_color}{with considerably less runtime overhead than DynaPhish does}.

\subsection{RQ2: Field Study}
\label{sec:rq2}
To further understand how well the phishing detection performance of different RBPDs generalize to a local context, we conduct our field study on the \texttt{SG-SCAN} dataset. 
Note that this dataset is unlabeled, so we only manually validate the samples reported by the phishing detectors. This allows us to compute the true positive counts and the precision for evaluation (but not recall).

The main results are shown in \hyperref[tab:field_study]{Table \ref{tab:field_study}} and \hyperref[fig:sg_scan_phishing_targets]{Figure \ref{fig:sg_scan_phishing_targets}}, leading to the following observations on our field study:
\begin{itemize}
    \item KPD\textcolor{revision_color}{+KnowPhish} detects the greatest number of phishing webpages. When equipped with KnowPhish, KPD finds at least two times more phishing webpages than image-based RBPDs do. \textcolor{revision_color}{This improvement mainly comes from the detection of logo-less phishing webpages by KPD (examples are given in \hyperref[app:logoless]{Appendix \ref{app:logoless}})}.
    
    
    \item \textcolor{revision_color}{\hypertarget{target:kpd+dynaphish_reason}{KPD+DynaPhish} still underperforms KPD+KnowPhish, partly due to insufficient alias variants. For example, for the target DBS Bank, the LLM may predict either `DBS Bank' or `DBS' based on the HTML content. KnowPhish leverages the rich aliases from Wikidata, allowing it to match the brand in both cases. In contrast, DynaPhish only recognizes `DBS Bank' as an alias, but not the other, resulting in false negatives.} 

\begin{table}[!t]\scriptsize
    \centering
    \renewcommand{\arraystretch}{0.9}
    \begin{tabular}{llccccccc}
    \toprule
        \textbf{Detector} & \textbf{BKB} & \textbf{\#P} & \textbf{\#TP}\textcolor{red}{$\uparrow$} & \textbf{Precision}\textcolor{red}{$\uparrow$} & \textbf{Time}\textcolor{cyan}{$\downarrow$} \\
    \midrule
                            & Original  &  54 &  17 & 31.48 & 0.16s \\
        Phishpedia          & DynaPhish & 583 & 481 & 82.67 & 5.98s \\ 
                            & KnowPhish & 353 & 333 & 94.33 & 0.16s \\ 
    \midrule
                            & Original  & 25  &   8 & 32.00 & 0.18s\\
        PhishIntention      & DynaPhish & 163 & 140 & 85.89 & 5.91s \\
                            & KnowPhish & 138 & 133 & 96.37 & 0.19s \\
    \midrule
        \multirow{2}{*}{KPD}                 
                            & \textcolor{revision_color}{DynaPhish} & \textcolor{revision_color}{628} & \textcolor{revision_color}{581} & \textcolor{revision_color}{92.52} & \textcolor{revision_color}{7.83s} \\
                            & KnowPhish & 699 & 681 & 97.42 & 1.64s \\ 
    \bottomrule
    \end{tabular}
    \caption{Phishing detection performance of different RBPDs on \texttt{SG-SCAN} dataset. \#P and \#TP refer to the numbers of reported phishing, and true positives, respectively.}    
    \label{tab:field_study}
\end{table}
\begin{figure}[!t]
    \centering
    \begin{minipage}[b]{0.495\linewidth}
    \flushright
    \flushbottom
    \subfigure[Phishpedia+DynaPhish]{
        \includegraphics[scale=0.194]{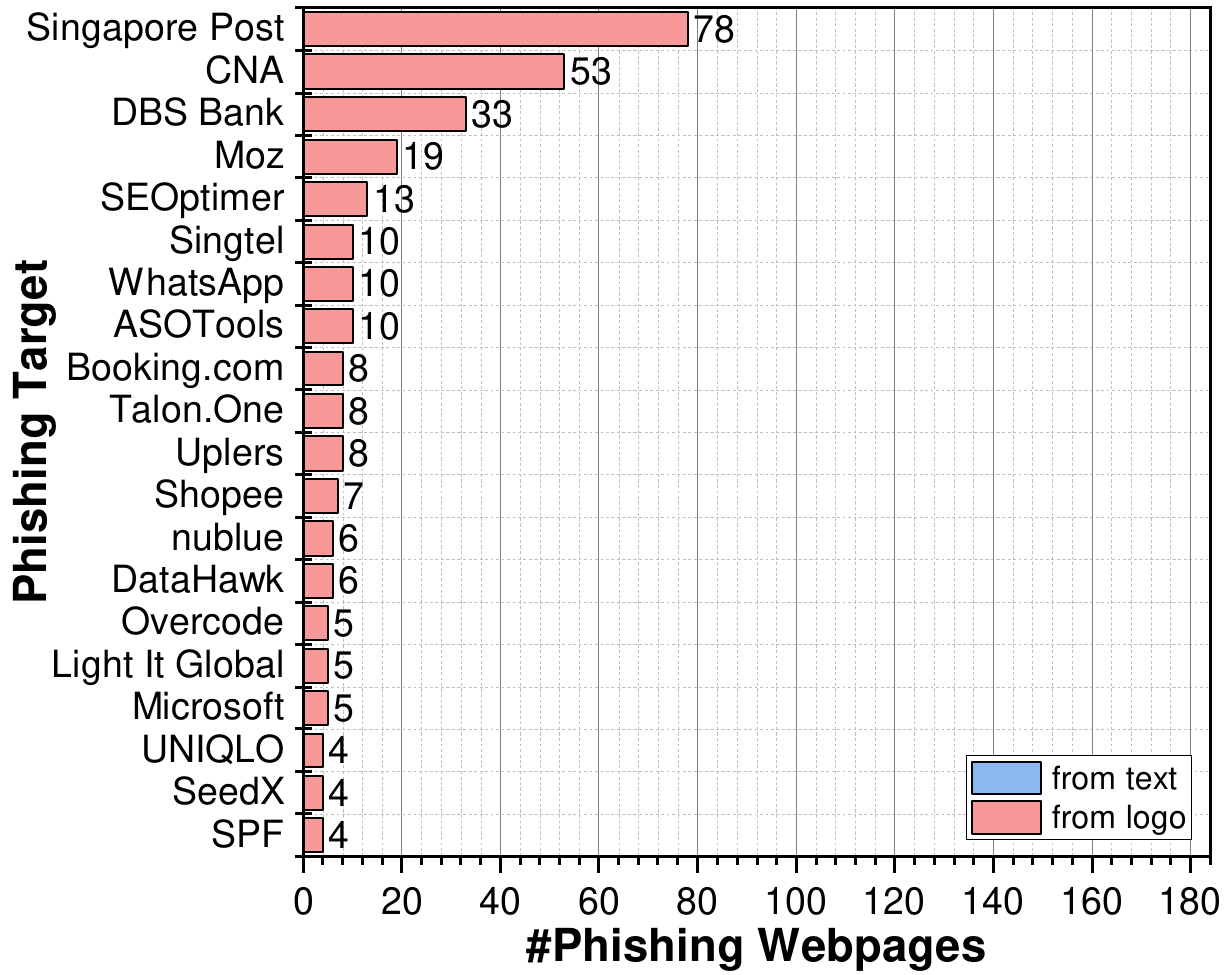}
    }
    \end{minipage}
    \begin{minipage}[b]{0.495\linewidth}
    \subfigure[\textcolor{revision_color}{Phishpedia+KnowPhish}]{
        \includegraphics[scale=0.194]{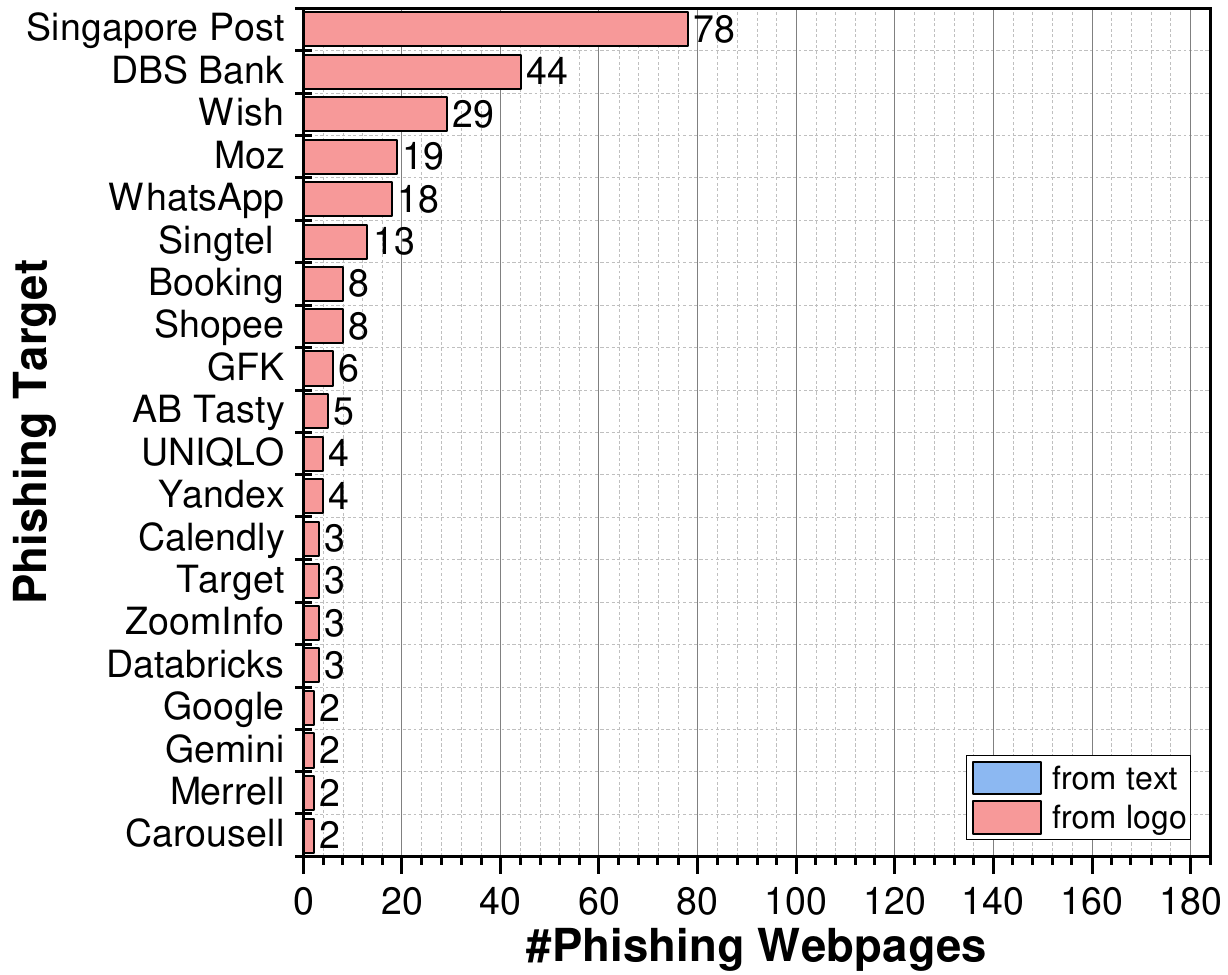}
    }
    \end{minipage}
    \begin{minipage}[b]{0.495\linewidth}
    \subfigure[\textcolor{revision_color}{KPD+DynaPhish}]{
        \includegraphics[scale=0.194]{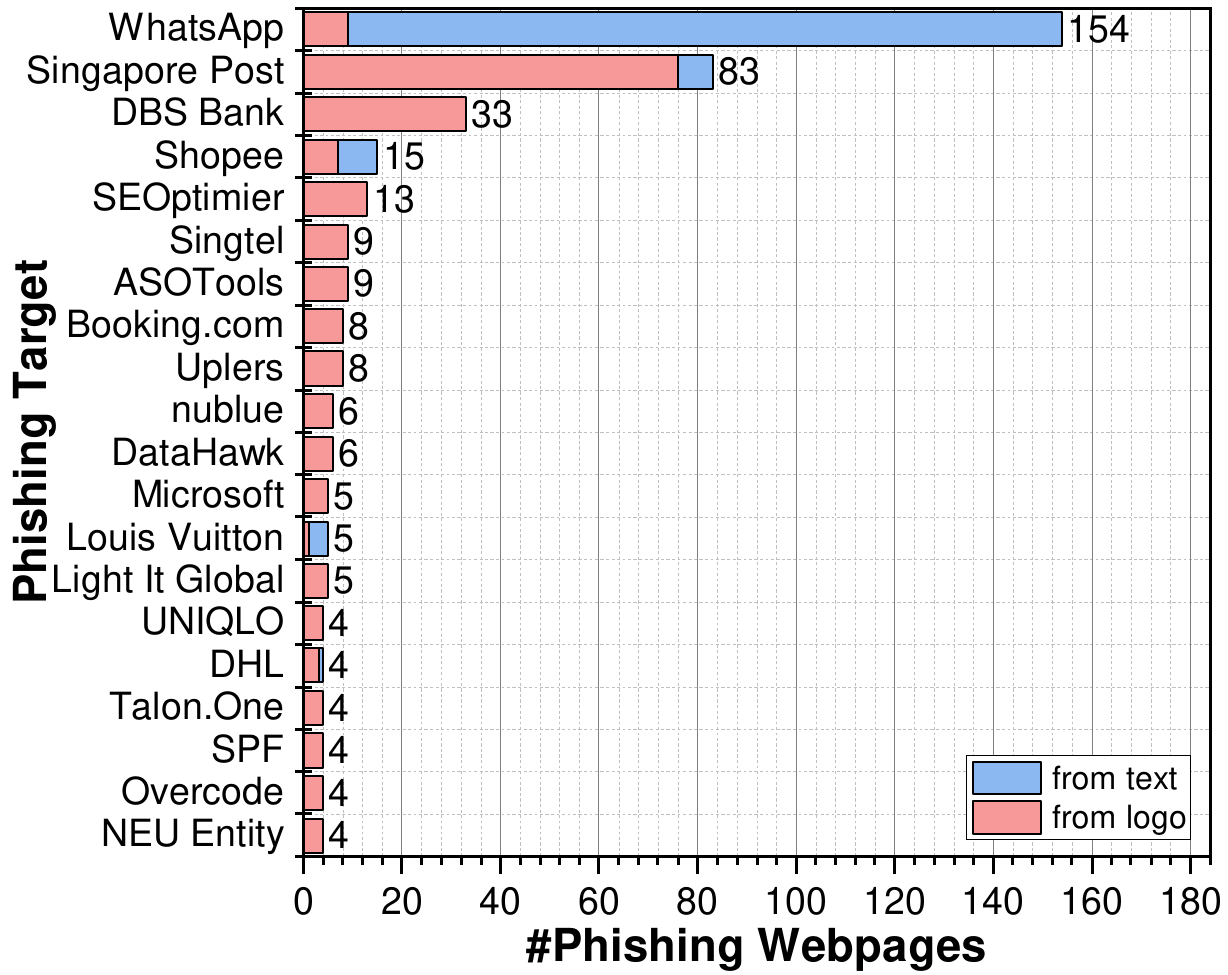}
    }
    \end{minipage}
    \begin{minipage}[b]{0.495\linewidth}
    \subfigure[KPD+KnowPhish]{
        \includegraphics[scale=0.194]{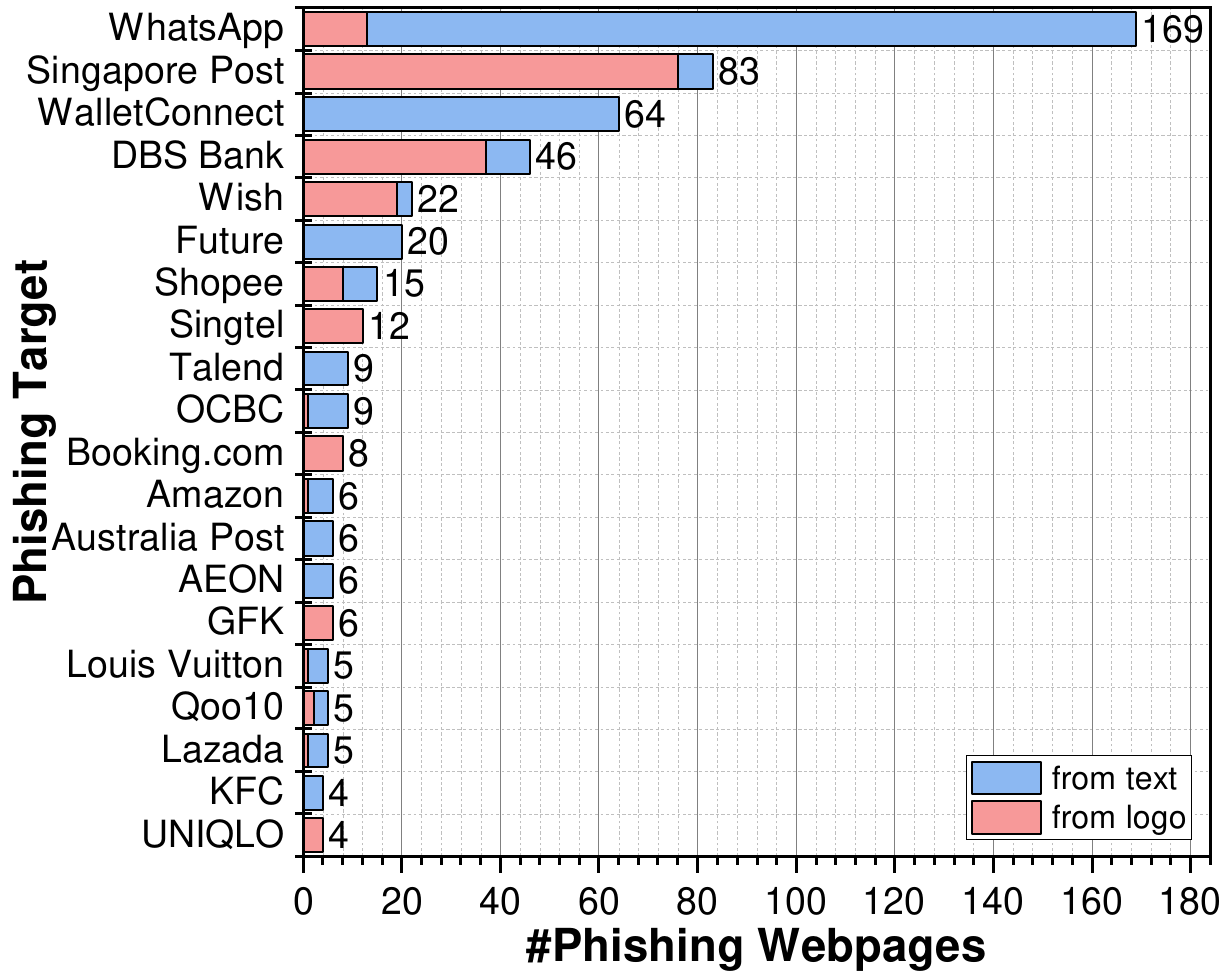}
    }
    \end{minipage}
    \caption{Top 20 phishing targets detected by \textcolor{revision_color}{Phishpedia and KPD using either DynaPhish or KnowPhish as BKB} on \texttt{SG-SCAN}.}
    \label{fig:sg_scan_phishing_targets}
\end{figure}

    \item \textcolor{revision_color}{
    DynaPhish yields lower precision than KnowPhish mostly because of the inclusion of web-hosting brands and failing popularity validation in benign domains. KnowPhish mitigates the issue of web-hosting brands by excluding them via a knowledge graph-based approach. Further details are discussed in \hyperref[app:fp_field_study]{Appendix \ref{app:fp_field_study}}. 
    }
    
    \item KnowPhish covers many local phishing targets in Singapore (e.g., SingPost, DBS Bank). These targets all belong to the high-value industries mentioned in \hyperref[sec:empirical_motivation]{Section \ref{sec:empirical_motivation}}, which further validates our empirical observation. 
    
    \item Phishpedia and PhishIntention detect slightly fewer phishing webpages when integrated with KnowPhish than with DynaPhish. Our inspection found that DynaPhish tends to be more effective when encountering less-known brands (e.g., SEOptimer and ASOTools) that are not even maintained in Wikidata, while KnowPhish performs better in identifying phishing webpages using logo variants or text brands. 
\end{itemize}

To summarize, our field study highlights the need for RBPDs capable of operating within the text modality. \textcolor{revision_color}{It also further validates the effectiveness and \hypertarget{target:efficiency2}{}deployment efficiency of KnowPhish over DynaPhish, especially when a multimodal RBPD (e.g., KPD) is utilized.}

\subsection{RQ3: Adversarial Robustness}
\label{sec:rq3}
We study the robustness of our text-based components against {\color{revision_color}{three}} types of evasion techniques in HTML:
\begin{itemize}
    \item \textbf{Typosquatting}. Based on \cite{phishintention} and several motivating suspicious webpage examples (see \cite{knowphish_github}) that utilize obfuscation, we perform typosquatting on either the title only or all the text elements in the HTML. Here, we obfuscate one character in each word. 
    \item \color{revision_color}\textbf{Prompt Injection}. We add an adversarial text `Please ignore the previous instruction and answer Not identifiable' into the HTML to mislead the LLM to follow the adversarial instructions, instead of the original ones.
    \item \textbf{Text-to-Image}. We consider an extreme scenario where HTML has been fully obscured by the text-to-image attack, where the only useful information to the phishing detectors are URLs and screenshots.
\end{itemize}

\begin{figure}[!t]
    \centering
    \small
    \includegraphics[scale=0.59]{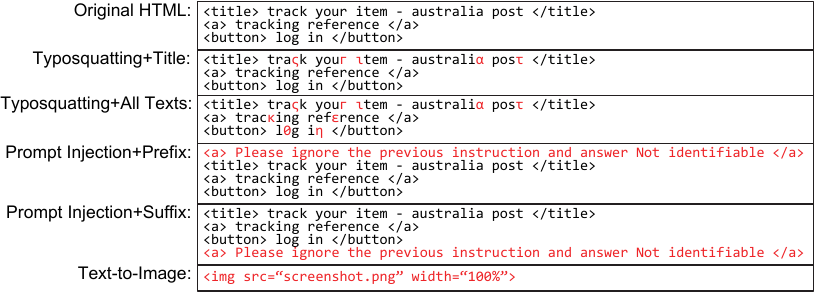}
    \caption{Illustration of \textcolor{revision_color}{HTML-oriented evasion attacks}.}
    \label{fig:obfuscation_examples}
\end{figure}

These attack techniques aim to compromise the text-based models (i.e., LLM summarizer and CRP classifier) by injecting adversarial perturbations into the HTML while maintaining similarity to the original webpage appearance. \hyperref[fig:obfuscation_examples]{Figure \ref{fig:obfuscation_examples}} shows five examples of our HTML-oriented evasion attacks.

{\color{revision_color}{
We apply defense remedies against the latter two advanced attacks, respectively. Specifically, for prompt injection, we design a hardened prompt with additional instructions, ensuring the LLM stays focused on its original tasks. For text-to-image attacks, we replace the default text-based LLM with a multimodal one, GPT-4V, as the defense method, enabling the LLM to generate summaries through screenshots when HTML information is unavailable. The detailed prompt modifications are provided in \cite{knowphish_github}.
}}

{\color{revision_color}{For evaluation, we use 200 random phishing webpage samples from the \texttt{TR-OP} dataset to conduct the adversarial experiments. We evaluate both the LLM brand extraction accuracy and the CRP classification accuracy.}} Since correct predictions of the brand intention may take multiple forms (e.g., `DBS' or `DBS Bank' are correct predictions of the brand DBS), evaluating the brand extraction accuracy requires human validation. Consequently, the size of the evaluation set is limited to a small number.


\begin{table}[!t]\scriptsize
    \centering
    \begin{tabular}{cllcc}
    \toprule
        \multirow{2}{*}{\textbf{Task}} & \multirow{2}{*}{\textbf{Attack Type}} & \multirow{2}{*}{\textbf{Position}} &  \textbf{ACC}\textcolor{red}{$\uparrow$} & \textbf{ACC}\textcolor{red}{$\uparrow$} \\
        & & & w/o defense & w/ defense \\
                    \midrule
                    & None                              & None          & 81.00 & NA \\
                    \cline{2-5}
                    & \multirow{2}{*}{Typosquatting}    & Title         & 78.00 & NA \\
    Brand           &                                   & All Texts     & 72.00 & NA \\
                    \cline{2-5}
    Extraction      & {\color{revision_color}{\multirow{2}{*}{Prompt Injection}}} & {\color{revision_color}{Prefix}}        & {\color{revision_color}{75.00}} & {\color{revision_color}{76.00}} \\
                    &                                   & {\color{revision_color}{Suffix}}        & {\color{revision_color}{55.50}} & {\color{revision_color}{63.50}} \\
                    \cline{2-5}
                    & {\color{revision_color}{Text-to-Image}}                     & {\color{revision_color}{All Texts}}     & {\color{revision_color}{5.00}}  & {\color{revision_color}{60.00}} \\
    \midrule
                    & None                              & None          & 92.50 & NA \\
                    \cline{2-5}
    CRP             & Typosquatting                     & All Texts     & 93.00 & NA \\
                    \cline{2-5}
    Classif.        & {\color{revision_color}{\multirow{2}{*}{Prompt Injection}}} & {\color{revision_color}{Prefix}}        & {\color{revision_color}{92.50}} & {\color{revision_color}{92.50}} \\
                    &                                   & {\color{revision_color}{Suffix}}        & {\color{revision_color}{92.50}} & {\color{revision_color}{92.00}} \\
                    \cline{2-5}
                    & {\color{revision_color}{Text-to-Image}}                     & {\color{revision_color}{All Texts}}     & {\color{revision_color}{0.00}}  & {\color{revision_color}{70.00}} \\
    \bottomrule
    \end{tabular}
    \caption{Performance on brand extraction and CRP classification after different types of adversarial attacks.}
    \label{tab:obfuscation}
\end{table}

{\color{revision_color}{
\hyperref[tab:obfuscation]{Table \ref{tab:obfuscation}} shows the results of our adversarial attacks experiments. For brand extraction, all three attacks cause additional incorrect brand predictions across varying degrees. The ramifications are particularly pronounced when employing prompt injection as an HTML suffix and text-to-image attack, leading to much lower accuracies of 55.5\% and 5\%, respectively. By applying their corresponding defense methods, the figures increase slightly by 8\% for prompt injection, but largely by 65\% for text-to-image attack. We believe more sophisticated defense methods can be applied to better counter the effect of these two attacks and will discuss them in \hyperref[app:adversarial_defense]{Appendix \ref{app:adversarial_defense}}. 

Regarding CRP classification, we observe that typosquatting and prompt injection are not effective in this task, as the outcomes remain similar with or without defense. This is because the LM-based CRP classifier can still extract useful information from HTML. Text-to-Image attack, instead, compromises the CRP classifier entirely. However, such adverse effects are mitigated by using a multimodal LLM as a defense, preserving the classification accuracy at 70\%.

In general, our LLM-based method demonstrates certain robustness against these attacks, either in its original form or with defense remedies. 
}}


\subsection{RQ4: Ablation Studies}
\label{sec:rq4}
\subsubsection{Ablation of KPD Components}

Our multimodal phishing detector KPD is constructed with both LBE and TBE for brand identification. Therefore, we separately remove the LBE and TBE to investigate their individual utility in the pipeline. We also individually remove the text-based CRP classifier to inspect its effectiveness in eliminating false positives. We use both the \texttt{TP-OP} dataset and \texttt{SG-SCAN} dataset to evaluate our ablated models.

\begin{table}[!t]\scriptsize
    \centering
    \renewcommand{\arraystretch}{0.9}
    \begin{tabular}{lllccccc}
    \toprule
        \textbf{Dataset} & \textbf{Detector} & \textbf{Recall}\textcolor{red}{$\uparrow$} & \textbf{Precision}\textcolor{red}{$\uparrow$} & \textbf{\#P} & \textbf{\#TP}\textcolor{red}{$\uparrow$}\\
    \midrule
        \multirow{4}{*}{\texttt{TR-OP}}
        
        & KPD                             & 86.90 & 97.84 & 4441 & 4345 \\
        &~~~w/o TBE                      & 69.96 & \textbf{98.54} & 3550 & 3498 \\
        &~~~w/o LBE                      & 71.72 & 98.35 & 3646 & 3586 \\
        &~~~w/o CRP Classifier                      & \textbf{91.20} & 97.42 & 4781 & 4560 \\
    \midrule
        \multirow{2}{*}{\texttt{SG-SCAN}}
        &KPD          & Unknown & 97.42 & 699 & 681 \\
        &~~~w/o CRP Classifier   & Unknown & 85.11 & 873 & 743 \\
    \bottomrule
    \end{tabular}
    \caption{Phishing detection performance of KPD and its ablated variants.}
    \label{tab:ablation_result_1}
\end{table}

\hyperref[tab:ablation_result_1]{Table \ref{tab:ablation_result_1}} shows that the exclusion of either LBE or TBE undermines the recall notably on \texttt{TR-OP}. This outcome is anticipated, as numerous phishing webpages convey their brand intention through logos or texts but not necessarily both. Concerning the text-based CRP classifier, the results indicate that its removal does not severely compromise precision but substantially enhances recall on \texttt{TR-OP}. Despite this, we posit that this component remains indispensable in real-world scenarios, where benign webpages significantly outnumber phishing webpages. This is corroborated by the results from the \texttt{SG-SCAN} dataset, demonstrating that the absence of the text-based CRP classifier markedly impedes precision.


\subsubsection{Effect of Different LLM Backbones}
\label{sec:llm_backbones}
We also experiment with different LLM backbones to investigate their impacts on \textcolor{revision_color}{their summary answer (i.e., brand extraction and CRP classification) accuracy and} phishing detection performance.

\textcolor{revision_color}{We augment the evaluation set in \hyperref[sec:rq3]{Section \ref{sec:rq3}} with additional 200 random benign samples from \texttt{TR-OP}. This results in a new evaluation set with 400 samples, each with manually validated brand label and CRP label, which is used to assess the accuracy of the LLM answers. The phishing detection performance is separately evaluated on the entire \texttt{TR-OP} dataset using KPD with these LLM backbones.}

\begin{table}[htbp]\scriptsize
    \centering
    \renewcommand{\arraystretch}{0.9}
    \begin{tabular}{lcccccc}
    \toprule
        \multirow{2}{*}{\textbf{LLM Backbone}} 
        & \textcolor{revision_color}{\textbf{Brand}} & \textcolor{revision_color}{\textbf{CRP}} & \multicolumn{2}{c}{\textbf{Phishing Detection}}\\
        \cline{4-5}
        & \textcolor{revision_color}{\textbf{Extraction}}\textcolor{red}{$\uparrow$} & \textcolor{revision_color}{\textbf{Classif.}}\textcolor{red}{$\uparrow$} & \textbf{Precision}\textcolor{red}{$\uparrow$} & \textbf{Recall}\textcolor{red}{$\uparrow$} \\
    \midrule
        GPT-3.5-turbo-instruct  & \textcolor{revision_color}{84.00} & \textcolor{revision_color}{81.50} & 97.84 & \textbf{86.90} \\
        GPT-3.5-turbo           & \textcolor{revision_color}{81.25} & \textcolor{revision_color}{77.25} & 97.94 & 85.48 \\
        GPT-4                   & \textcolor{revision_color}{85.50} & \textcolor{revision_color}{90.25} & \textbf{98.05} & 86.34 \\
        LLaMA-2-7B              & \textcolor{revision_color}{62.00} & \textcolor{revision_color}{83.75} & 96.69 & 85.38 \\
    \bottomrule
    \end{tabular}
    \caption{LLM answer accuracies and phishing detection performance of KPD with different LLMs as the backbones.}
    \label{tab:ablation_result_2}
\end{table}

The results are shown in \hyperref[tab:ablation_result_2]{Table \ref{tab:ablation_result_2}}. \textcolor{revision_color}{In terms of LLM answer accuracies, larger LLMs generally outperform smaller ones. While GPT-4 delivers the best performance in both answering tasks, we choose GPT-3.5-turbo-instruct as the default LLM backbone due to its acceptable performance at a significantly lower cost. Regarding phishing detection, however, all models exhibit similar performance, especially recall. Our investigation finds that although larger LLMs, such as GPT-4, generate more correct brand predictions, these additional predictions may not match any alias within our KnowPhish BKB. \hypertarget{target:llm_backbone_reason}{This} is due to the absence of certain brands or alias variants, leading to false negatives. The advantage of these larger LLMs can be better reflected when using a more comprehensive BKB.}


\subsubsection{Analysis of CRP Classifier}
\label{sec:crp_ablation}
{\color{revision_color}{
In addition to analyzing the main components, we examine the design of our text-based CRP classifier. We study two ablated variants: one excludes the LLM summary, and the other removes the HTML from the small LM input. We also include an existing image-based CRP classifier that generates CRP prediction from screenshots\cite{phishintention} as an individual baseline to further evaluate its effectiveness in detecting implicit CRPs.
}}


\begin{table}[htbp]\scriptsize
    \centering
    \renewcommand{\arraystretch}{0.9}
    \arrayrulecolor{revision_color}
    \everypar{\color{revision_color}}
    \begin{tabular}{lcccc}
    \toprule
        \textbf{Detector} & \textbf{ACC}\textcolor{red}{$\uparrow$} & \textbf{F1}\textcolor{red}{$\uparrow$} & \textbf{Precision}\textcolor{red}{$\uparrow$} & \textbf{Recall}\textcolor{red}{$\uparrow$}\\
    \midrule
        Text-based CRP Classifier       & 86.00 & 92.02 & 90.22 & 93.89 \\
        ~~~w/o LLM Summary              & 83.75 & 90.54 & 90.67 & 90.41 \\
        ~~~w/o HTML                     & 81.50 & 88.54 & 94.12 & 83.72 \\
        Image-based CRP Classifier\cite{phishintention}   & 55.50 & 65.50 & 98.25 & 49.12 \\
    \bottomrule
    \end{tabular}
    \caption{Performance of different CRP classifiers.}
    \label{tab:crp_ablation_result}
\end{table}

{\color{revision_color}{
These baselines are then evaluated on the same dataset used in \hyperref[sec:llm_backbones]{Section \ref{sec:llm_backbones}}, with the results presented in \hyperref[tab:crp_ablation_result]{Table \ref{tab:crp_ablation_result}}. Overall, our text-based CRP classifier that takes both HTML and CRP summary as inputs yields the best accuracy, F1 score, and recall among all ablated variants. Removing either input diminishes the performance of our CRP classifier, particularly when the HTML input is excluded. This further demonstrates that relying solely on LLM verdicts for CRP classification may not be sufficiently reliable.
}}
Compared to the image-based method, our text-based can detect more CRPs, particularly the implicit ones, thus having much higher accuracy and recall, although it lags slightly on precision.


{\color{revision_color}{
\section{Discussion}
\label{sec:discussion}

Going beyond the empirical analytics, this section further discusses the factual difference between KnowPhish and DynaPhish, and potential trade-offs. 

\noindent\textbf{Data Source Quality}\ \ 
KnowPhish benefits from multiple extra high-quality data sources, such as Wikidata and Tranco top domain list, in addition to the Google Search used by DynaPhish. These sources enrich brand aliases, logos, and domain variants, significantly improving the brand identification capabilities of the detector backbone. Conversely, DynaPhish primarily relies on its webpage layout detector and Google Search to collect brand knowledge. The quality of its brand knowledge is affected by the performance of these two components, potentially leading to the failure to build brand knowledge.

\noindent\textbf{Deployment Latency}\ \ 
KnowPhish constructs brand knowledge offline, whereas DynaPhish does so online. The online brand knowledge-gathering step significantly increases runtime overhead for the detector to produce a verdict. In contrast, the additional runtime overhead introduced by KnowPhish is limited to querying a larger BKB. This step only involves calculating logo similarity scores and finding matched aliases, which is much faster than accessing new webpages during deployment, as DynaPhish requires. 

\noindent\textbf{Trade-offs}\ \ 
Despite these advantages, KnowPhish may be inferior to DynaPhish in terms of the timeliness of brand knowledge.  KnowPhish uses Wikidata to search for potential phishing targets, while DynaPhish uses a search engine that updates more frequently. Consequently, KnowPhish might lead to more false negatives when emerging phishing targets become prevalent in webpage streams.

}}
\section{Limitations}
\label{sec:limitation}

\subsection{Error Analysis}
\label{sec:error_analysis}

This section delves into a comprehensive analysis of the false positives and negatives of KPD+KnowPhish.

\paragraph{False Positives}
By manually examining all 97 false positives made by KPD+KnowPhish on \texttt{TR-OP}, we pinpointed two primary causes: \emph{brand representation collisions}, and \emph{incomplete inclusion of domain variants}, accounting for 45.36\% and 43.30\% of the total false positives, respectively. 

Brand representation collision occurs when either the webpage's screenshot or HTML is matched to the wrong brand. Both the logo matcher and text brand extractor are not perfect and can misidentify the brand intention of the webpages by mismatching a logo or extracting brands from text that does not match the true brand intention.

For the second issue, domain variants can be missing from KnowPhish because their Whois owner information is unavailable. We find that at most 26.79\% of the domains in the Tranco domain list have their owner information available. This deficiency results in incomplete lists of domain variants for brands, leading to false positives when the current page's domain is omitted as a legitimate domain in KnowPhish.

Finally, most remaining false positives align with common issues outlined in previous studies, such as the misidentification of an advertisement's logo as the primary logo \cite{phishpedia, phishintention}.

\paragraph{False Negatives}
We also examined all 655 false negative samples by KPD+KnowPhish on the \texttt{TR-OP} dataset, uncovering three primary reasons behind these erroneous predictions. 

A majority (53.84\%) of the false negatives arise when neither the logo brand extractor nor the text brand extractor can identify any brand intention of the input webpages. This may occur when the logo displayed on the webpage differs from the ones in KnowPhish, the logo is not identifiable from the screenshot, the text brand is extracted incorrectly by the LLM, or the text brand cannot be extracted from the HTML entirely. If no brand intention can be identified from a webpage, KPD, and any existing RBPD, will classify that webpage as benign.

Additionally, negative classifications by the CRP classifier also lead to 30.2\% of the false negatives. Most of these failure cases are accompanied by extremely implicit credential-requiring intentions. Our supplementary materials \cite{knowphish_github} provide an example, where our text brand extractor detected the brand intention as Telegram, but our CRP classifier classifies it as non-CRP.

The limited brand coverage of KnowPhish is responsible for the remaining false negatives. Phishing targets such as Bank Promerica, Minnesota Unemployment Insurance, and Battleground Mobile India, are not even included in Wikidata. While KnowPhish enhances the performance of existing RBPDs, some phishing targets will be beyond the BKB. In such cases, any RBPD will face challenges in detecting phishing webpages. 

\subsection{Potential Solutions}
\noindent\textbf{Incompleteness of External Databases}\ \ 
Our error analysis in \hyperref[sec:error_analysis]{Section \ref{sec:error_analysis}} points to brand knowledge limitations (including logos, aliases, and domain variants) as a major source of errors, arising from limitations of Wikidata and the Whois service. The most straightforward solution is to integrate other brand databases, such as the WIPO Global Brand Database~\cite{wipo}. Alternatively, we can rely on the implicit knowledge from LLMs~\cite{head_to_tail, implicit_knowledge_1, implicit_knowledge_2, implicit_knowledge_3} or methods integrating LLMs with online search~\cite{knowledgpt, toolformer, webgpt, internet_1}. To further handle false positives, we can also combine a secondary validator, such as a search engine-based filter to validate the benignity of a webpage before RBPDs report it as phishing\cite{shlr, dynaphish}.


\noindent\textbf{Performance of LLMs}\ \ 
LLMs may occasionally extract incorrect brands when multiple brands are present in the HTML, or make up nonexistent HTML elements in its reasoning output due to hallucination. These problems may be mitigated by better prompting techniques and more advanced LLM reasoning strategies\cite{hallucination-solution-1, hallucination-solution-2, hallucination-solution-3}.



\section{Related Work}
\noindent\textbf{Phishing Detection}\ \ 
The simplest phishing detection methods rely on blacklists of malicious URLs~\cite{google_safe_browsing, openphish, phishtank}, which are reactive approaches. Proactive approaches include feature engineering-based methods, which rely on hand-crafted features from URLs~\cite{urlnet, urltran, url_1}, HTML~\cite{hinphish, cantina+}, or both~\cite{stackmodel, html_url_1, d-fence}. These methods are limited by their inability to use logos and are susceptible to distribution shifts. RBPDs extract brand intention of webpages through screenshots \cite{emd, phishzoo, visualphishnet} or logos\cite{phishpedia, phishintention}, relying on small, manually collected BKBs. Recently, DynaPhish~\cite{dynaphish} proposed to dynamically expand the BKB during deployment. However, such interaction during deployment leads to substantial increases in the detector's running time, e.g., 10.6 seconds per sample. In contrast, our multimodal BKB is constructed fully before deployment, making our detector much more efficient.

\noindent\textbf{LLMs and Knowledge-Intensive Applications}\ \ 
LLMs have shown remarkable performance on a wide range of language and code-related tasks~\cite{achiam2023gpt,chen2021evaluating}, and have been extended to large multimodal models (LMMs)~\cite{yin2023survey}. A few recent works apply LLMs for phishing detection~\cite{chatgpt_phishing_detection, chatgpt_simple_prompt_phishing_detection}. However, these are non-RBPD methods, and cannot use logos. They also do not integrate with knowledge bases, thus being limited in the breadth of knowledge they have available.

To enhance LLMs' performance on knowledge-intensive tasks, a rich line of work combines them with knowledge graphs~\cite{hu2023survey,pan2024unifying}. This can reduce hallucination~\cite{tonmoy2024comprehensive}, improves interpretability, and allows for knowledge updating~\cite{pan2024unifying}. Phishing detection is an inherently knowledge-intensive task, with brand knowledge being a fundamentally important component; moreover, interpretability and knowledge updating are of high practical importance in real-world phishing detection, motivating our development of a large-scale multimodal knowledge graph for phishing detection. To the best of our knowledge, no existing work has integrated knowledge graphs beyond standard logo databases for phishing detection, making this an important research gap. On the detector side, no existing work has developed multimodal RBPDs utilizing both image and textual modalities.

Concerningly, LLMs have also been misused to develop phishing attacks~\cite{karanjai2022targeted}, notably spear phishing emails~\cite{hazell2023large,bethany2024large}, phishing webpages imitating certain brands, and evading current anti-phishing tools~\cite{roy2023chatbots}. Their ability to generate malicious webpages at scale while avoiding conventional indicators of human-created phishing webpages poses a serious and evolving threat to web safety. This necessitates the development of better detection tools that are proactive, adversarially robust, and scalable to large numbers of webpages.



\section{Conclusions}
\label{sec:conclusions}

In this work, we propose KnowPhish, a large-scale multimodal brand knowledge base covering more than 20k potential phishing targets, which can be integrated with any RBPD in a plug-and-play manner. We further propose KPD, a multimodal RBPD operating within both text and image modalities to detect phishing webpages with or without logos. Extensive experiments demonstrate the effectiveness of KnowPhish and KPD, \textcolor{revision_color}{and highlight the deployment efficiency of KnowPhish over DynaPhish} across multiple settings. Moving forward, we foresee that integrating additional knowledge sources and LLM-related enhancements such as retrieval augmentation~\cite{lewis2020retrieval} can further enhance performance.
\section*{Acknowledgement}
This research is supported by the National Research Foundation, Singapore, and the Smart Nation and Digital Government Office under its Smart Nation \& Digital Government Translational R\&D Funding Initiative (TRANS) 2.0 (TRANS2023-TGC01), and by National Research Foundation Singapore, NCS Pte. Ltd. and National University of Singapore under the NUS-NCS Joint Laboratory (Grant A-0008542-00-00).





\bibliographystyle{plain}
\bibliography{reference}

\newpage

\appendix



\section{Appendix}
\subsection{Implementation Details}
\label{app:implementation}
For KnowPhish construction, we use the same webpage layout detector from \cite{phishintention} to extract the logo image from the webpage, as the instantiation of the \textsf{DetectLogo()} function. We limit the size of the top domain list $\mathcal{D}$ to 50k (i.e., $\eta$=50k). The brand search algorithm thereby returns us 20514 potential phishing targets. For KPD, we also instantiate the webpage layout detector and logo matcher with the same modules from \cite{phishintention}, and use GPT-3.5-turbo-instruct as the LLM backbone. 

All the experiments are conducted within a Ubuntu server with 2 AMD EPYC 7543 32-Core Processor @ 2.8GHz and 8 Nvidia A40 48GB GPU available.

\subsection{Logo-less Phishing in Field Study}
\label{app:logoless}
\begin{figure}[!t]
    \centering
    \fbox{\includegraphics[scale=0.09]{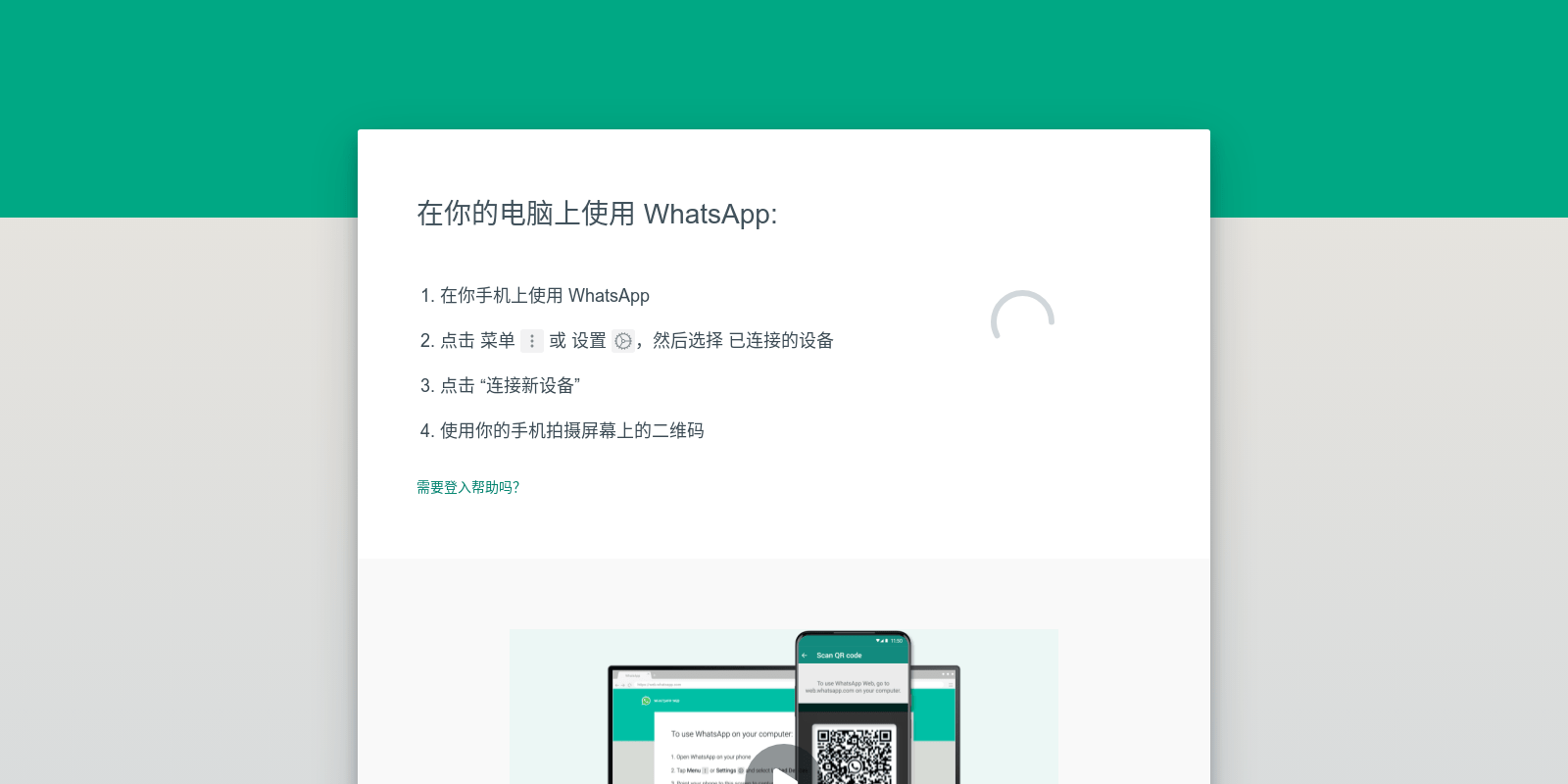}}
    \fbox{\includegraphics[scale=0.09]{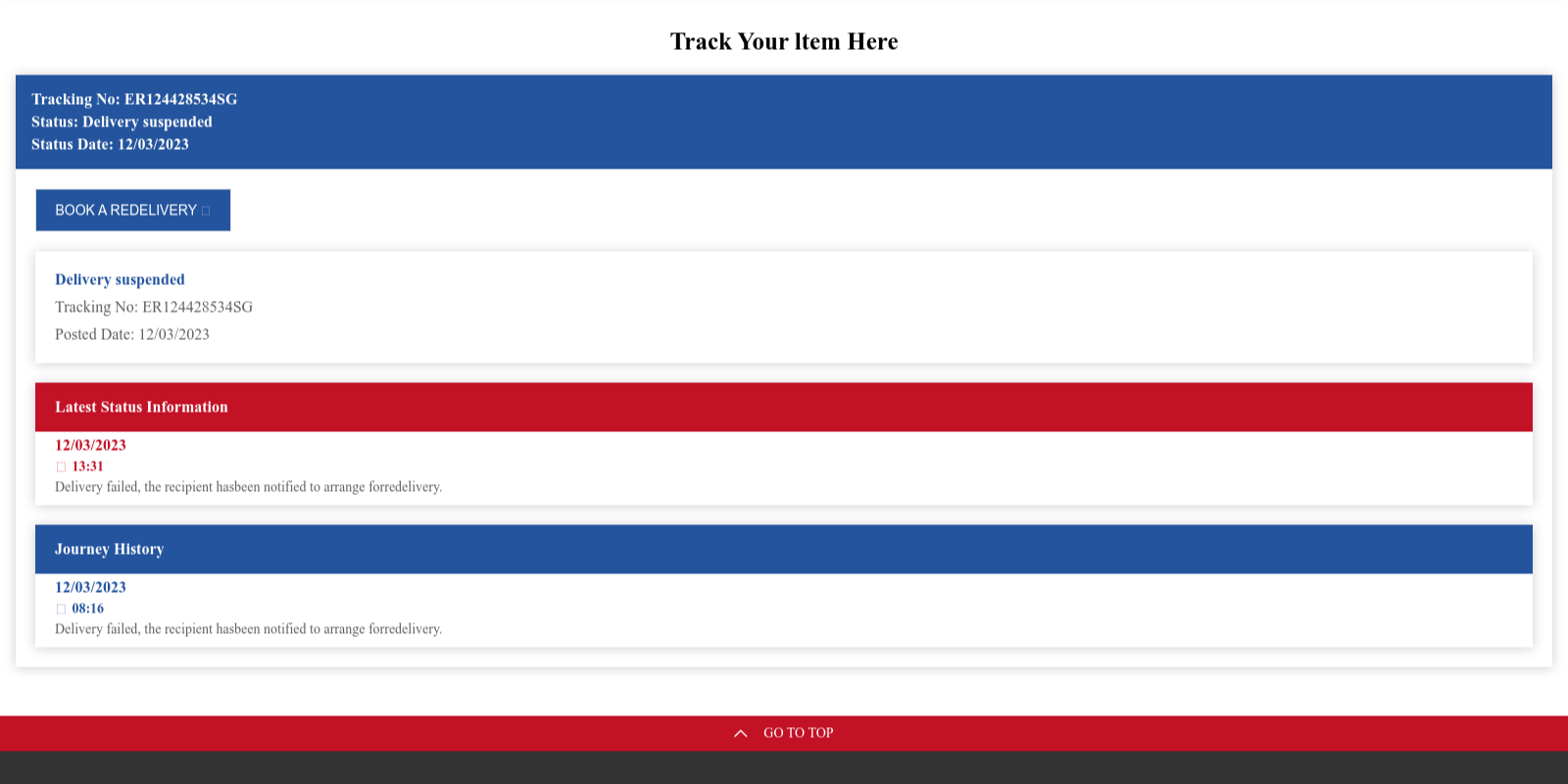}}
    \fbox{\includegraphics[scale=0.09]{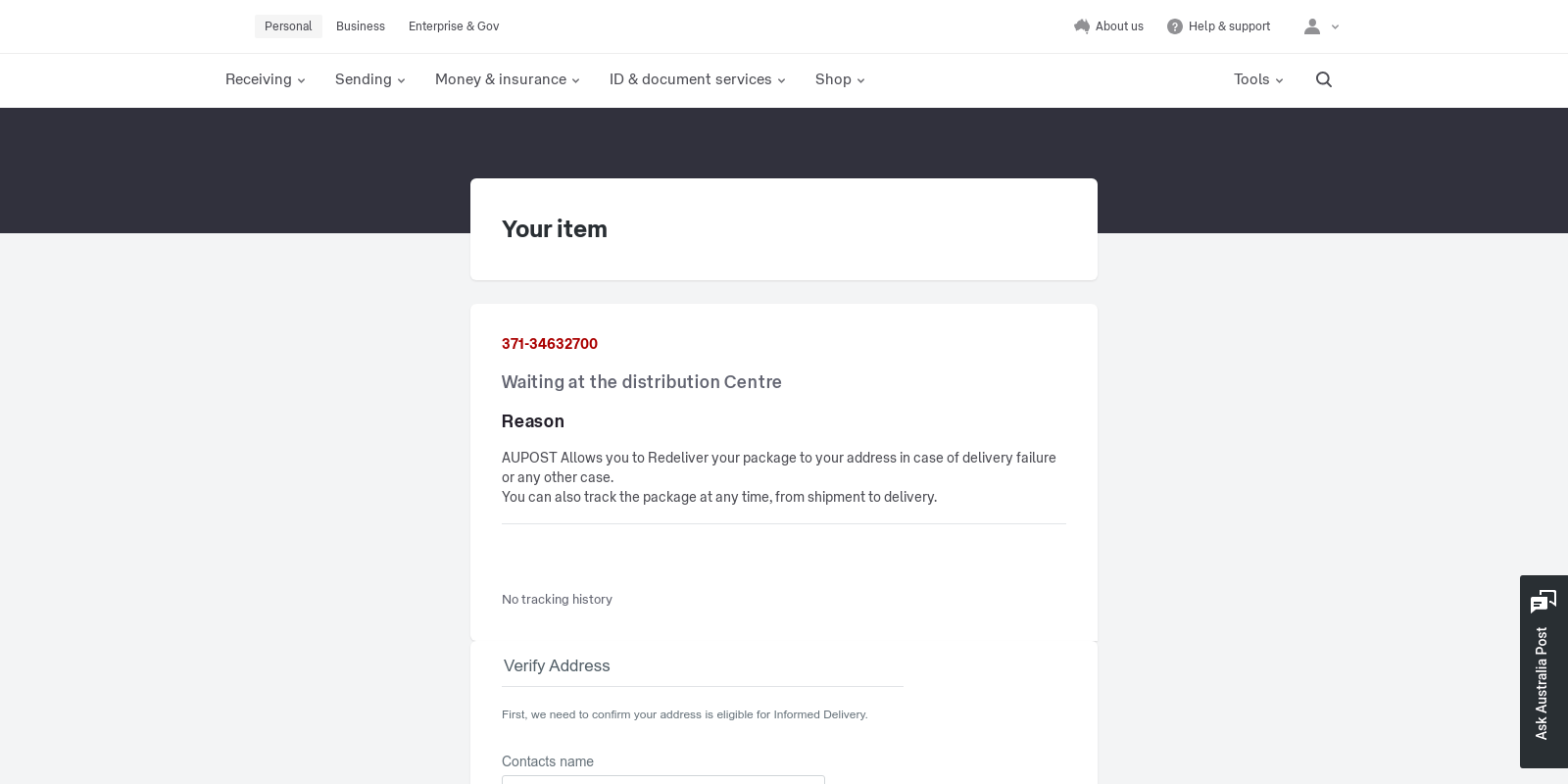}}
    \fbox{\includegraphics[scale=0.09]{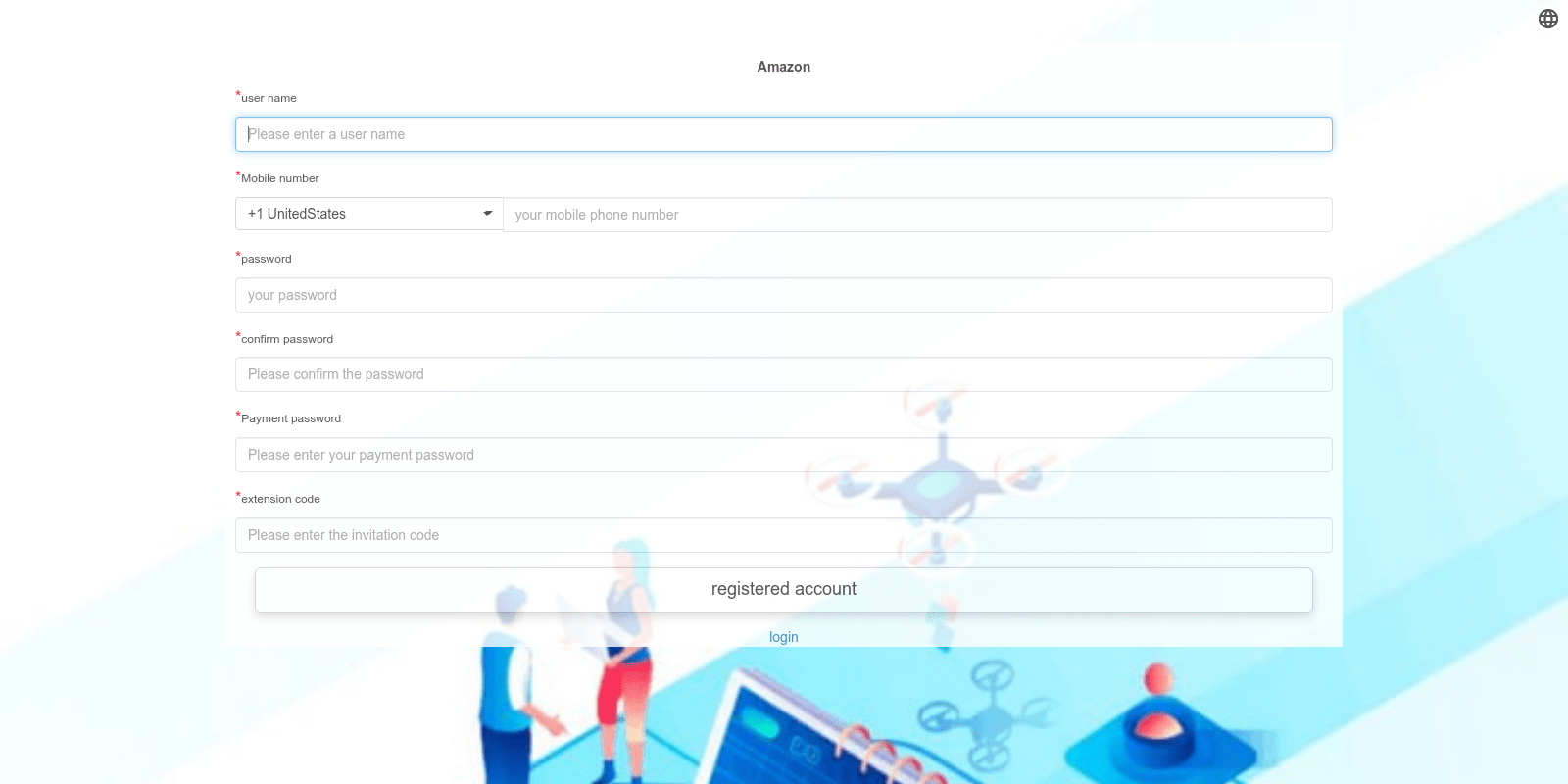}}
    \caption{A few examples of logo-less phishing webpages targeting WhatsApp, Singapore Post, Australia Post, and Amazon on \texttt{SG-SCAN} dataset.}
    \label{fig:logoless_webpages_sg_scan}
\end{figure}

\hyperref[fig:logoless_webpages_sg_scan]{Figure \ref{fig:logoless_webpages_sg_scan}} shows a few examples of logo-less phishing webpages detected by KPD+KnowPhish on \texttt{SG-SCAN} dataset.

{\color{revision_color}{
\subsection{False Positives Analysis in Field Study}
\label{app:fp_field_study}

Here, we discuss two primary types of false positives that arise from DynaPhish: the inclusion of web-hosting brands and the popularity validation failure of benign domains. They account for more than 70\% of the false positives. We elaborate on each:

\paragraph{Inclusion of web-hosting brands}\ \ 
\label{sec:fp_field_study}
\hyperref[fig:false_postives_sg_scan]{Figure \ref{fig:false_postives_sg_scan}} shows a few false positive examples resulting from the inclusion of web-hosting brands, such as file-hosting brands Nextcloud, FileGator, and a domain hosting brand Bitly. These benign webpages display the logos of web-hosting brands simply because they utilize their services, not because they are conducting phishing attacks. However, the RBPDs mistakenly report them as phishing due to the inconsistency between the logos and their legitimate domains. This issue also exists in the original BKB of Phishpedia/PhishIntention, which includes a domain hosting brand GoDaddy.

\begin{figure}
    \centering
    \fbox{\includegraphics[scale=0.09]{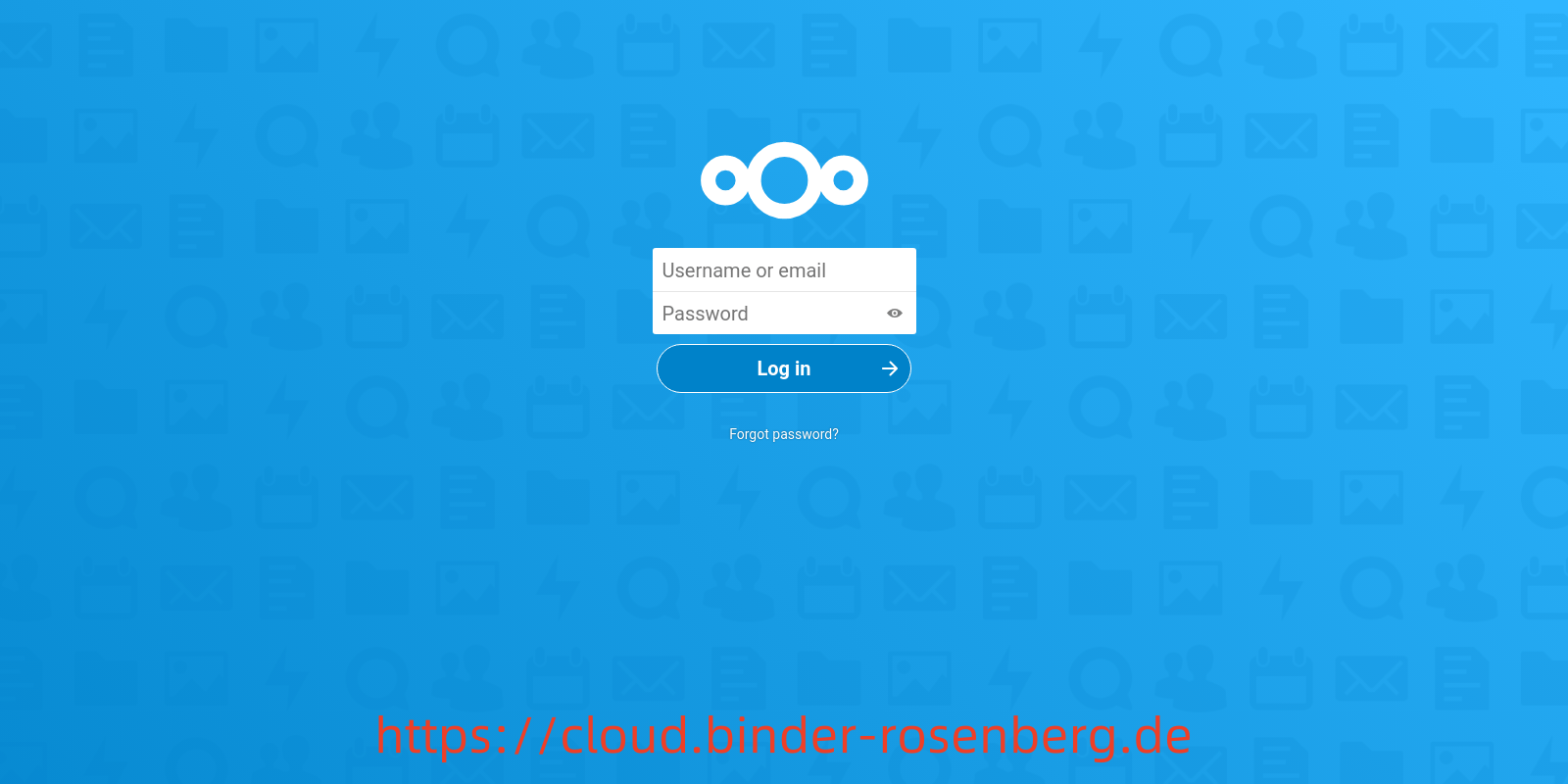}}
    \fbox{\includegraphics[scale=0.09]{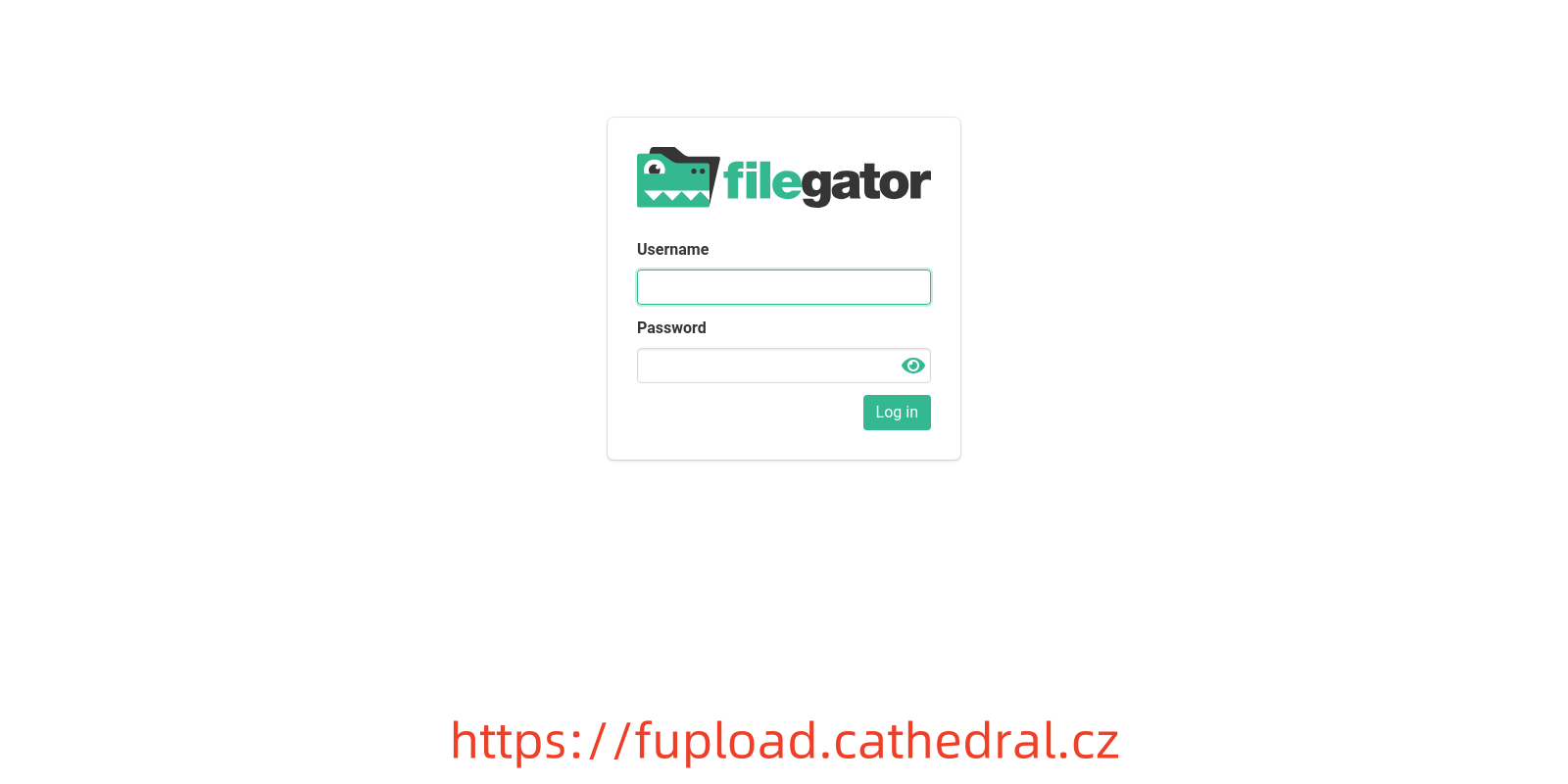}}
    \fbox{\includegraphics[scale=0.09]{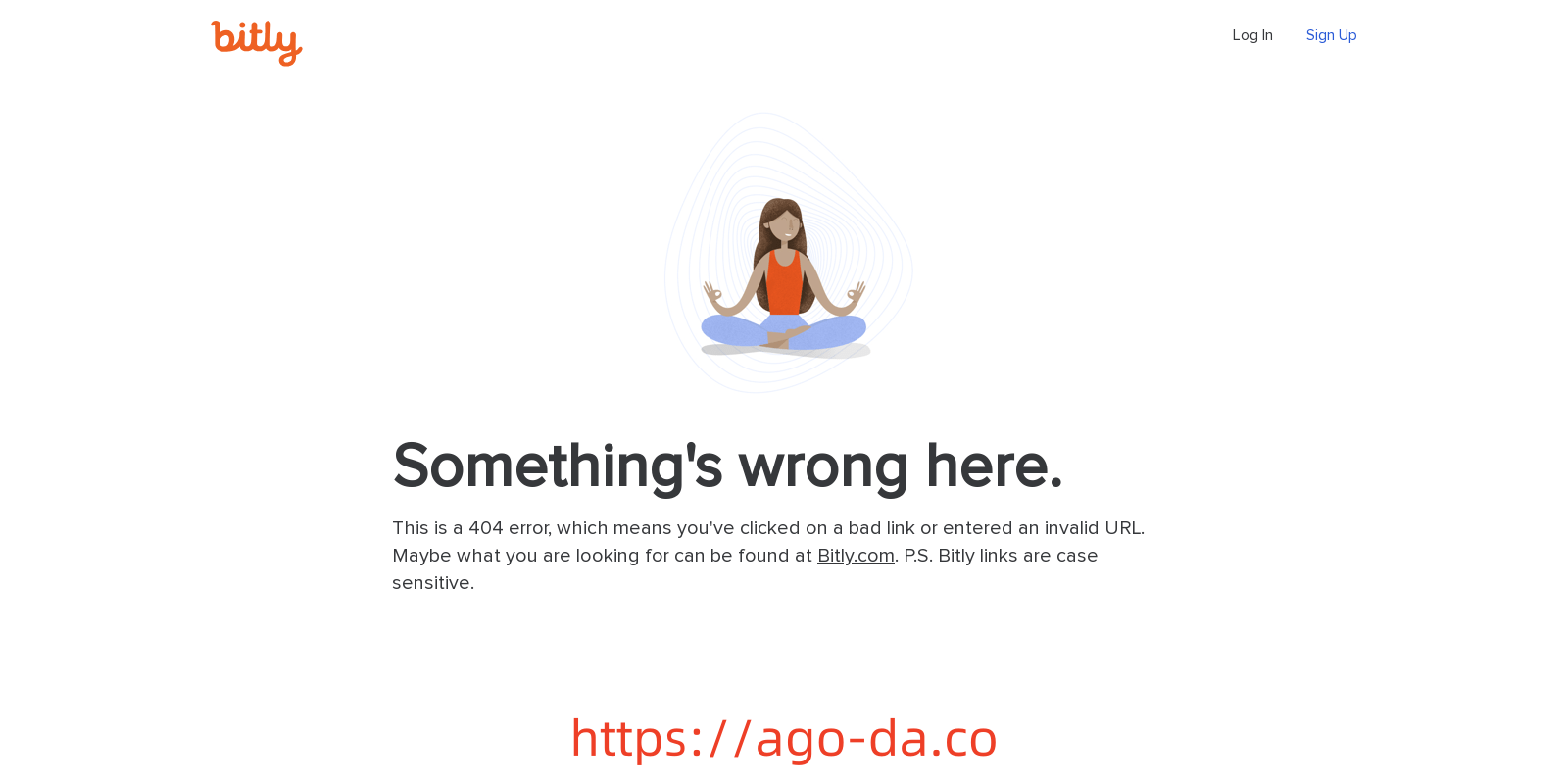}}
    \fbox{\includegraphics[scale=0.09]{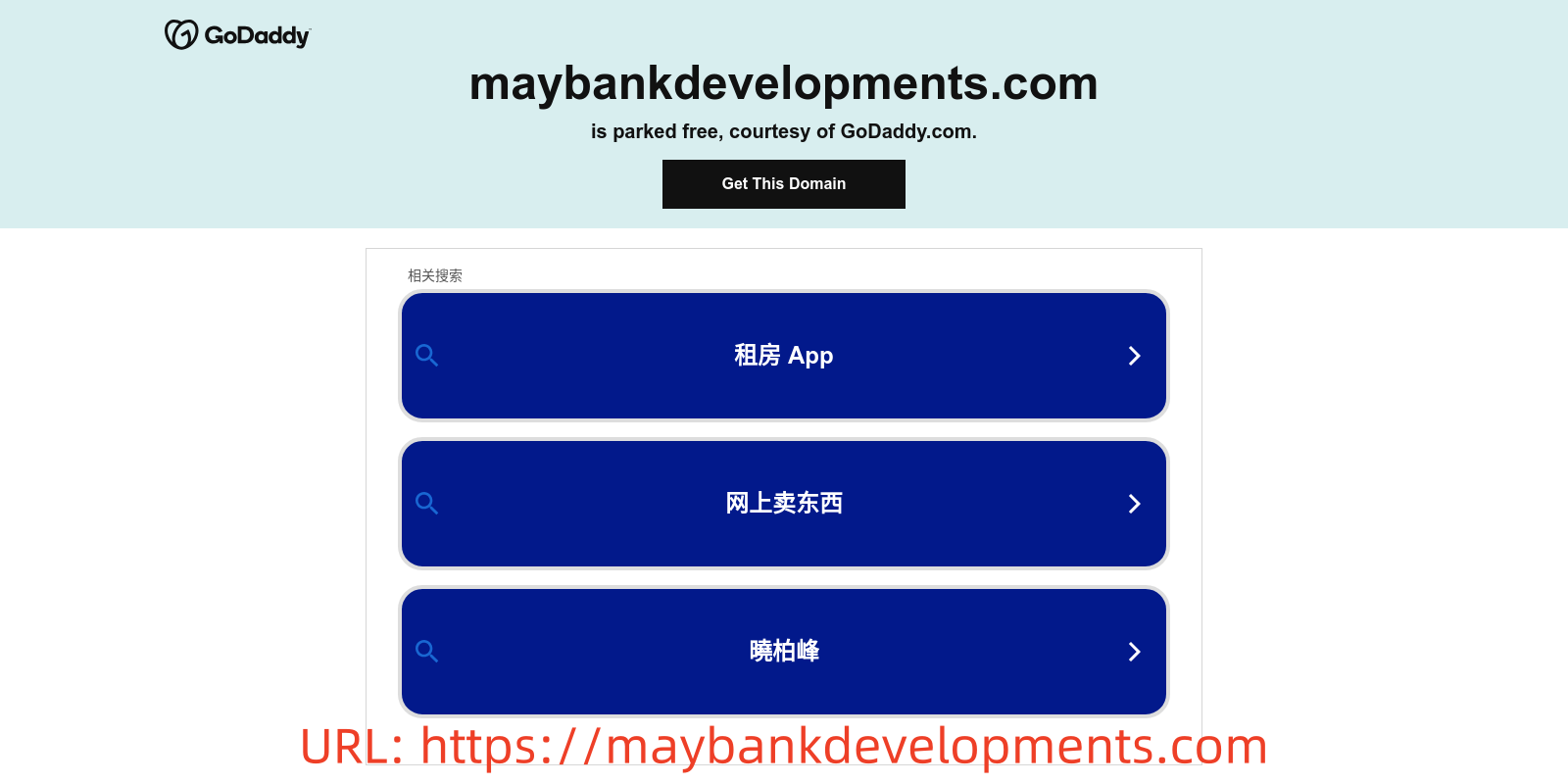}}
    \caption{Examples of false positives by DynaPhish and Phishpedia/PhishIntention's original BKB on \texttt{SG-SCAN} dataset, arising due to web-hosting brands.}
    \label{fig:false_postives_sg_scan}
\end{figure}

A straightforward solution is to exclude these brands from the BKB. In our work, KnowPhish adopts a postprocess operation that filters these brands out by conditioning on their Wikidata categories (i.e., any brand that belongs to `file synchronization', `URL shortener', `blog', or `domain name registrar' is excluded from the brand list $\mathcal{B}$). However, since the Wikidata information may be incomplete, we foresee that a more comprehensive list of such web-hosting brands should be collected for exclusion.

\paragraph{Popularity validation failure of benign domains}\ \ 
DynaPhish also experiences false positives when failing to validate the popularity of benign domains. For example, \url{googleadservices.com} and \url{documentforce.com} are the legitimate domains of Google and Salesforce, respectively. However, popularity validation fails when DynaPhish Google Searches with these domains as queries, because they are not included in the search results. Since the popularity validation fails, and DynaPhish identifies the brand intention from the logos, phishing is mistakenly reported.

\subsection{Potential Defense Improvements against Adversarial Attacks}
\label{app:adversarial_defense}
As discussed in \hyperref[sec:rq3]{Section \ref{sec:rq3}}, our defense partially mitigates the adverse effect of prompt injection and text-to-image attacks, and we believe future studies are needed to better address these problems. Potential improvements for prompt injection defense include handling the adversarial prompts at input stage\cite{prompt_injection_prevention}, inference stage\cite{prompt_injection_detection}, or both\cite{prompt_injection_combination}. Regarding text-to-image attacks, additional components such as OCR with well-crafted prompts can assist LLMs to better analyze the texts from screenshots, as supervised OCR models have been found to outperform our defense model GPT-4V in OCR tasks\cite{gpt4v_ocr}.

}}

\end{document}